\documentclass[prl,twocolumn,showpacs,preprintnumbers,amsmath,amssymb]{revtex4-1}

\usepackage{graphicx}
\usepackage{ifthen}
\usepackage{ifpdf}
\usepackage{color}
\definecolor{red}{rgb}{1.,0.,0.}
\definecolor{blue}{rgb}{0.,0.,1.}

\usepackage{latexsym}
\usepackage{amssymb}
\usepackage{amsmath}
\usepackage{bbm}
\usepackage{bm}
\newcommand{\half}{\mbox{\small $\frac{1}{2}$}}

\newcommand{\eexp}{\mbox{e}^}

\newcommand{\RR}{{\bf R}}
\newcommand{\VV}{{\bf V}}
\newcommand{\AB}{{\bf A}}
\newcommand{\ab}{{\bf a}}
\newcommand{\BB}{{\bf B}}
\newcommand{\CC}{{\bf C}}

\newcommand{\beq}[1]{\begin{eqnarray}\ifthenelse{#1=-1}{\nonumber}
{\ifthenelse{#1=0}{}{\label{e#1}}}}
\newcommand{\eeq}{\end{eqnarray}}
\newcommand{\be}{\begin{equation}}
\newcommand{\ee}{\end{equation}}

\newcommand{\bea}{\begin{eqnarray}}
\newcommand{\eea}{\end{eqnarray}}

\newcommand{\hide}[1]{}

\begin{document}

\title{Transconducting transition for a dynamic boundary coupled to several Luttinger liquids}

\author{B. Horovitz}
\affiliation{ Department of Physics, Ben Gurion University,
Beer Sheva 84105 Israel}
\author{T. Giamarchi}
\affiliation{DPMC-MaNEP, University of Geneva, 24 Quai Ernest Ansermet, 1211 Geneva 4, Switzerland}
\author{P. Le Doussal}
\affiliation{
Laboratoire de Physique Th\'eorique de l'Ecole Normale Sup\'erieure,
PSL University \\
CNRS, Sorbonne Universit\'es,
24 rue Lhomond, 75231 Paris Cedex 05, France
}

\begin{abstract}
We study a dynamic boundary, e.g. a mobile impurity, coupled to $N$ independent Tomonaga-Luttinger liquids (TLLs) each
with interaction parameter K.
We demonstrate that for $N\geq 2$ there is a quantum phase transition at $K\geq\half$, where the TLL phases lock together at the particle position, resulting in a non-zero transconductance equal to $e^2/Nh$. The transition line terminates  for strong coupling at $K=1- \frac{1}{N}$, consistent with results at large $N$.  Another type of a dynamic boundary is a superconducting (or a Bose-Eninstein condensate) grain coupled to $N\geq 2$ TLLs, here the transition signals also the onset of a relevant Josephson coupling.
\end{abstract}

\maketitle

\begin{figure*}
\centering
\includegraphics [width=0.24 \textwidth ]{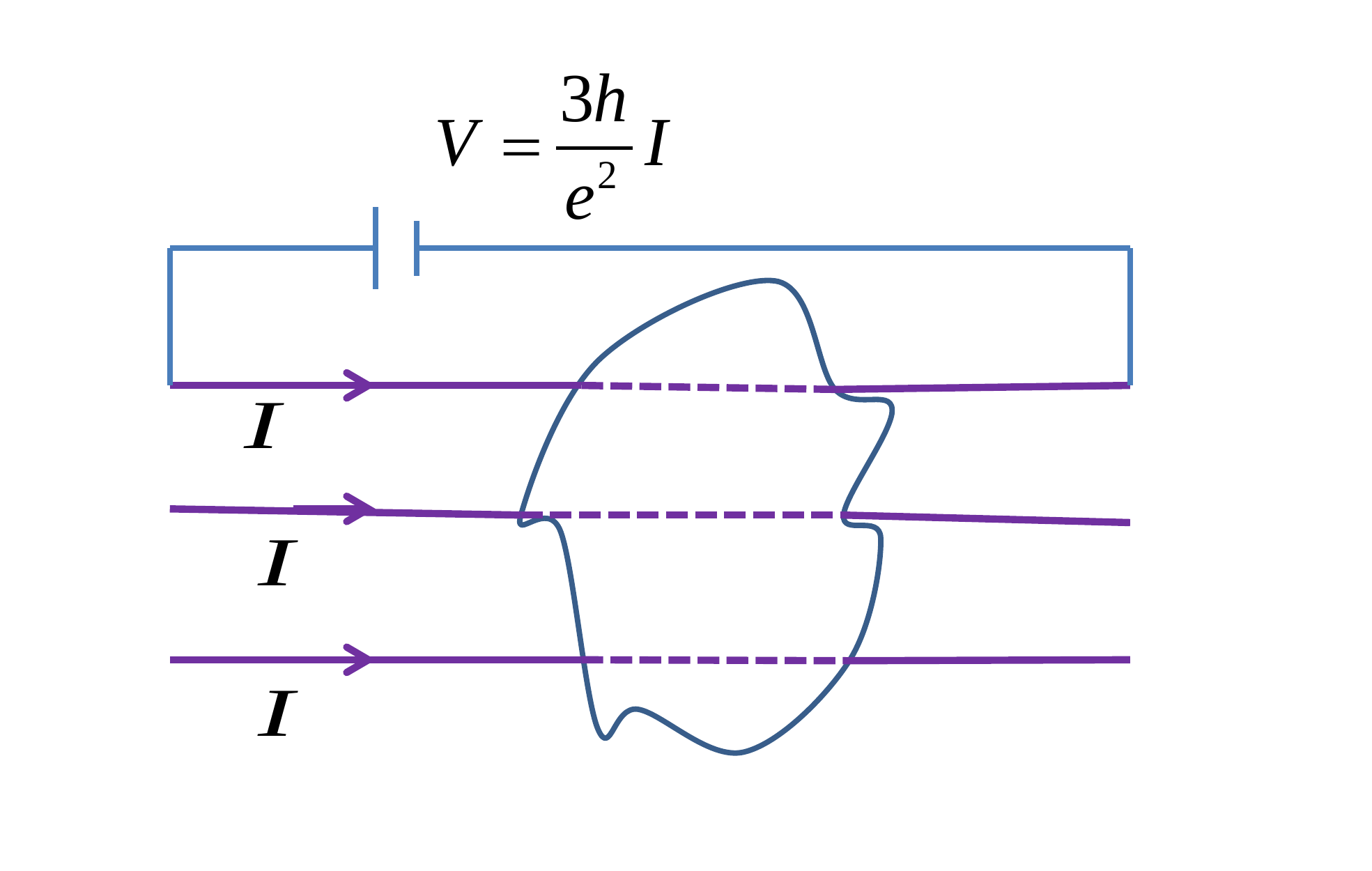}
\includegraphics [width=0.24 \textwidth ]{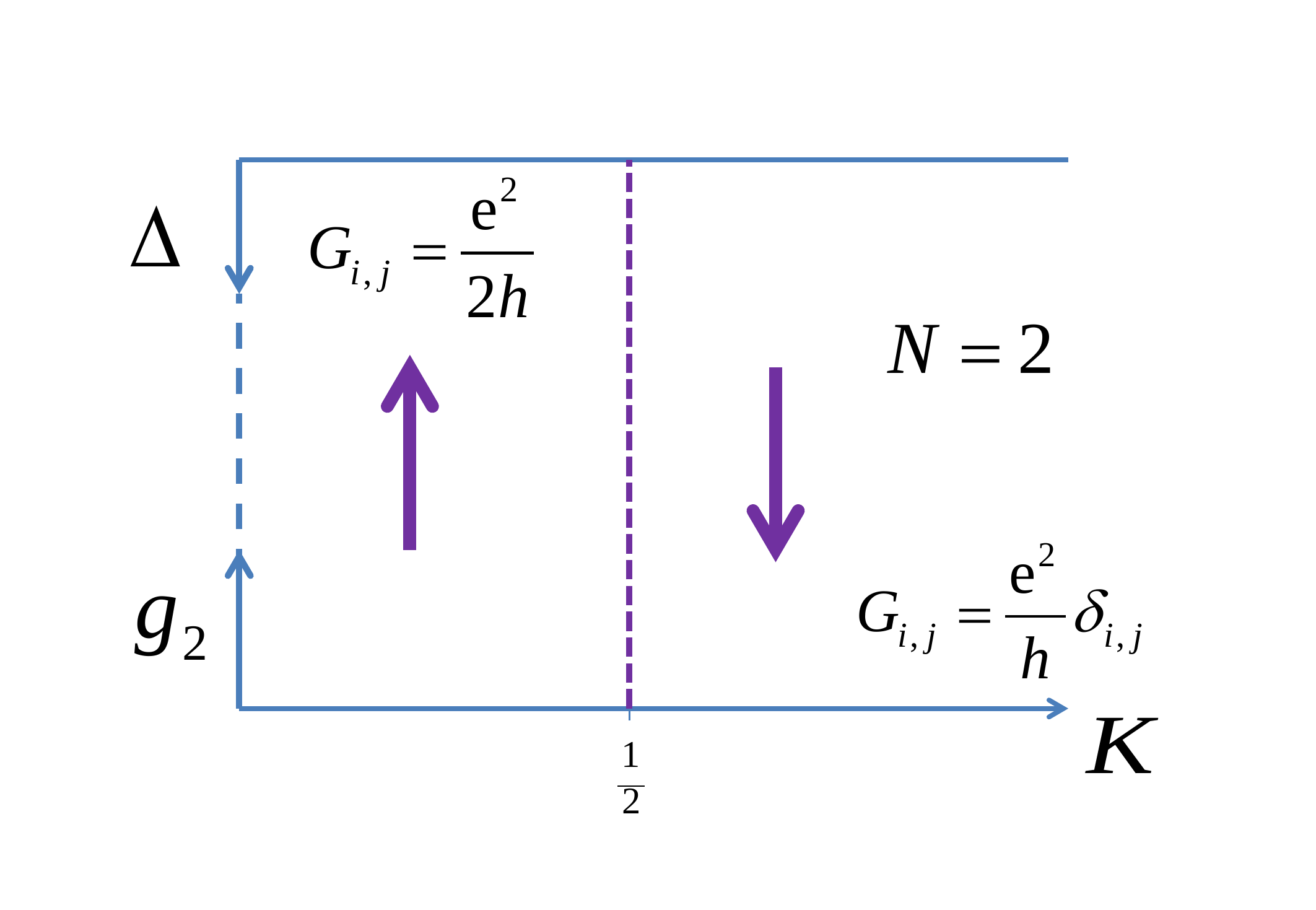}
\includegraphics [width=0.24 \textwidth ]{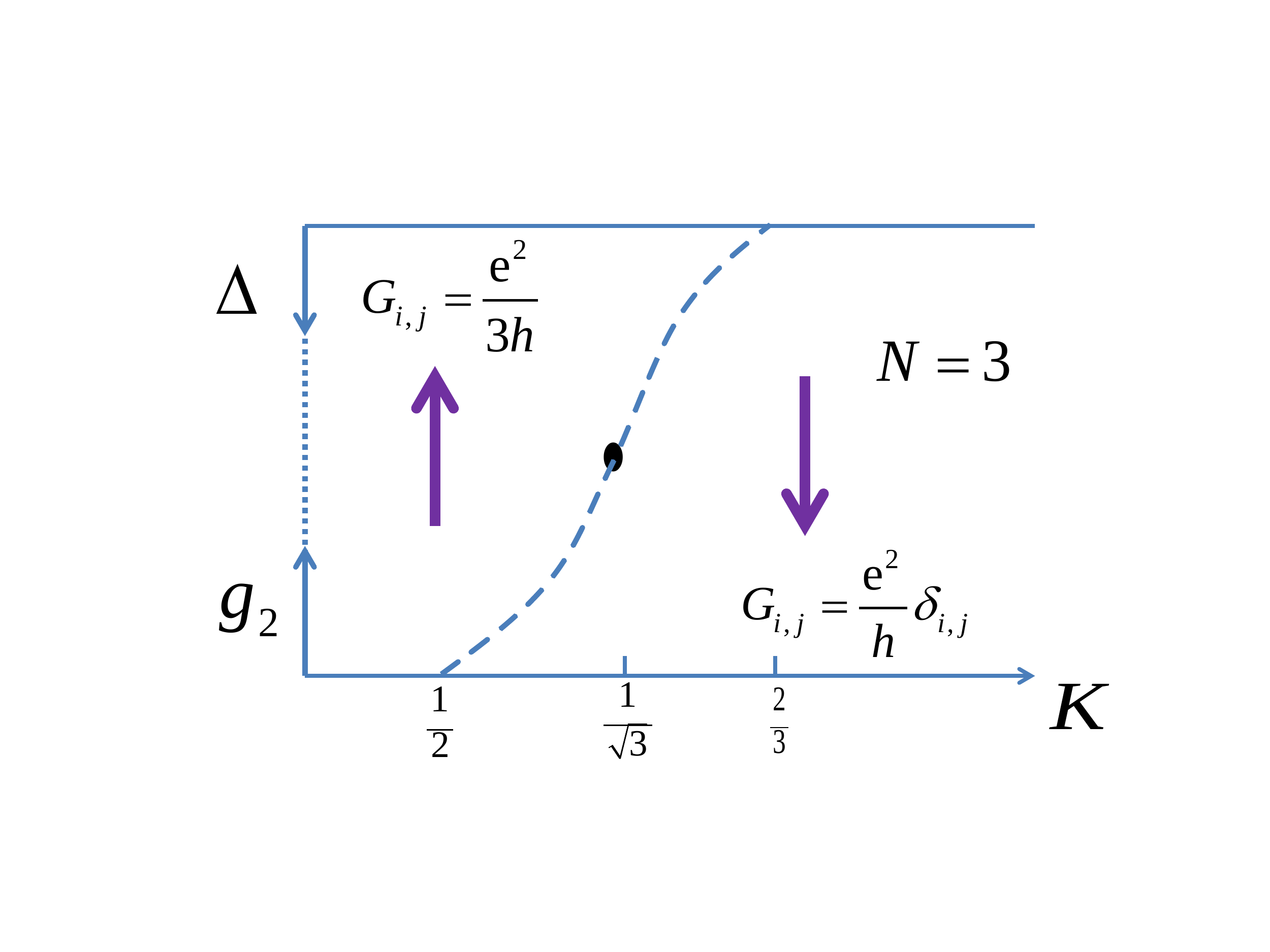}
\includegraphics [width=0.24 \textwidth ]{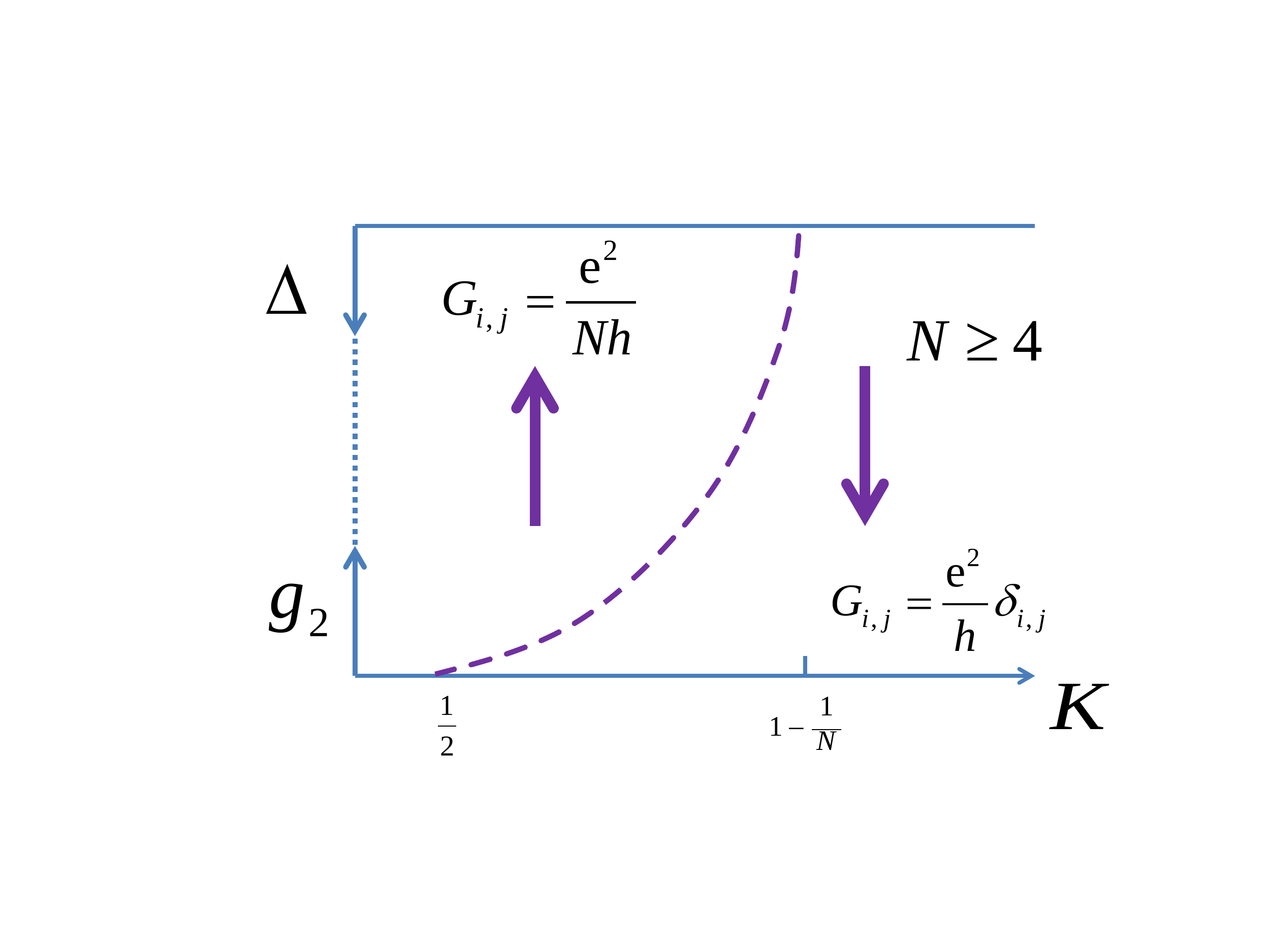}
\caption{ Setup and phase diagrams. The left frame shows a setup with a superconducting grain Josephson coupled to $N=3$ TLLs showing transconductance measurement in the coupled phase. The phase diagrams show $g_2$ or $\Delta$ ($\ln \Delta\sim -\sqrt{g_2}$ for large $g_2$) versus
 the TLL interaction parameter $K$ for $N=2,3,\geq 4$ respectively, in a mobile impurity system; for a superconducting grain, replace $K$ by $1/(2K_\rho)$ where $K_\rho$ is the TLL parameter in the charge sector; $K,K_\rho<1 (>1)$ correspond to repulsive (attractive) TLL interactions. Dashed lines are phase transition lines (lines of RG fixed points) separating a decoupled phase where $g_2\rightarrow 0$ is irrelevant and a coupled phase with strong transconductance where $g_2\rightarrow \infty$ (equivalently $\Delta\rightarrow 0$); RG flow directions are indicated by thick purple arrows (they are vertical since $K$ does not flow). The conductance matrices in the two phases are shown (add prefactor 2 for the spinfull superconducting grain). The conductance matrix varies continuously along the phase transition lines. The dot on the N=3 case corresponds to a self dual point at $K=1/\sqrt{3}$, see Eq. (\ref{e28}).}
\end{figure*}

 There is considerable interest in systems of $N$ independent one-dimensional Tomonaga-Luttinger liquids (TLLs) coupled via a dynamic boundary, e.g. a mobile impurity. This case has been realized in cold atom experiments employing a variety of impurity atoms, in either boson or fermion systems\cite{schirotzek,nascimbene,gadway,spethmann,zipkes,schmid,chikkatur,palzer,catani,fukuhara,meinert}.
These studies range from polaronic effects in bulk baths \cite{schirotzek,nascimbene,gadway}, the approach to equilibrium \cite{spethmann,zipkes,schmid,chikkatur} and more recently to one-dimensional cold atom gases \cite{palzer,catani,fukuhara,meinert}. Quantum impurities in TLL have been extensively studied \cite{zvonarev,schechter,massel,burovski,gamayun,kantian,andraschko,doggen,visuri}, focusing on the particle dynamics and response to an external force.

A second type of dynamic boundary is realized by a superconducting grain, or a Bose-Einstein condensate (BEC), illustrated in the left pannel of Fig. 1. This can be realized with wires formed on LaAlO$_3$/SrTiO$_3$ nanostructures \cite{stornaiuolo} or with carbon nanotubes. The latter system \cite{bouchiat} has in fact shown a surprisingly large values of supercurrents. Of further interest are topological superconductors with Majorana islands coupled to TLLs via multiterminals. Theoretical studies \cite{altland,sela,mora} show the phenomena of inter-terminal conductance, with possible realizations in various experimental setups \cite{marcus,kouwenhoven}. In dynamically coupled TLLs, as we show here, an analogous phenomenon, transconductance (see below), can occur even without Majorana states.

The case of an infinite number of TLLs was previously examined \cite{bh} and showed a phase transition in which
the impurity can localize for a repulsive TLL. Understanding the finite $N$ case
is thus important in view of the experimental realizations \cite{bouchiat,marcus,kouwenhoven} and the theoretical studies \cite{altland,sela,mora}. Further motivation for studying $N>1$ are studies of drag conductance in crossed carbon nanotubes \cite{flensberg,nazarov,klesse,ponomarenko,komnik,gao}.

In this Letter, using the renormalization group (RG) and duality methods akin to the study of quantum Brownian motion
in a periodic lattice \cite{yi1,yi2} we solve the problem of an impurity coupled to $N$ TLLs.
We demonstrate that there is a quantum phase transition for any $N\geq 2$
and show that the order parameter of this transition is provided by the transconductance, i.e. driving a current in one chain by applying a voltage on another chain. The transconductance vanishes in one phase and is $e^2/Nh$ in the strong coupled phase.
We show that the phase boundary interpolates between $K=\half$ at weak particle-TLL coupling and $K=1-\frac{1}{N}$ at strong coupling and
discuss the experimental consequences.

We start with the dynamic impurity problem, and eventually consider the equivalent superconducting grain system. We focus on equilibrium dynamics with imaginary time $\tau$.
The particle position is denoted by $X_\tau$ and its motion is described by the action
\beq{01}
&& S_{imp} = \int d\tau ~ \half M_0 \dot X_\tau^2 - g_0 \sum_{i=1}^N \rho_i(X_\tau,\tau)
\eeq
which includes the kinetic energy and identical instantaneous
contact interactions with the densities $\rho_i(x,\tau)$, $i=1,\cdots, N$ of the TLLs.
The density of each TLL entering in the coupling (\ref{e01}),
is related to its standard phase field \cite{giamarchi} $\phi_i(x,\tau)$ as
\beq{02}
\rho_i(x,\tau) = \rho_0+  \alpha_1 \rho_0\cos[2 \pi \rho_0 x - 2 \phi_i(x,\tau)]
\eeq
where $\rho_0$ is the average density of each TLL and $\alpha_1$ is non-universal.
Inserted in (\ref{e01}) it leads to a direct coupling of the particle position to
the oscillating part of the TLL density, which is the important effect here.
We have neglected in (\ref{e02}) higher harmonics of the density, as well
as the slowly varying part
$- \partial_x \phi_i(x,\tau)/\pi$ of the density, which leads to subdominant corrections to the particle motion at low frequencies.
The total action of the system is $S= S_{TLL} + S_{imp}$ where $S_{TLL}$ is the standard action
for $N$ independent TLLs with common Luttinger parameter $K$
\beq{03}
&& S_{TLL}=\sum_{i=1}^N \int dx d \tau \frac{1}{2 \pi K} [ (\partial_\tau \phi_i)^2 + (\partial_x \phi_i)^2 ]
\eeq
and we work in units such that the phonon velocity $u=1$. Let us express the position $X_\tau$ in
units of $(2\pi\rho_0)^{-1}$.
The action of the particle becomes
\beq{04}
S_{imp} =\half M\int_\omega \omega^2|X_\omega|^2-g\int_\tau \sum_{i=1}^N \cos[X_\tau-2\phi_i(X_\tau,\tau)] \nonumber\\
\eeq
where $M=M_0 (2 \pi \rho_0)^{-2}$, $g=\alpha_1 \rho_0 g_0$, and
here and below $\int_{\omega} f(\omega) \equiv \frac{1}{\beta} \sum_{\omega_n} f(\omega_n)
 \xrightarrow{\beta \to \infty}\int\frac{d\omega}{2\pi}$ with $\int_\tau= \int_0^{\beta} d\tau$. We focus on the
zero temperature $T=1/\beta  \to 0$ limit unless stated otherwise.

The theory defined by (\ref{e03}-\ref{e04}) is still highly non-linear in the particle position $X_\tau$ and
difficult to treat exactly. However, we claim that an equivalent long time theory is obtained by
\beq{05}
\cos[X_\tau-2\phi_i(X_\tau,\tau)]  \to \cos[X_\tau-2\phi_i(0,\tau)]
\eeq
This amounts to assuming that the TLL correlations are dominated by time differences $|\tau-\tau'|\gg |X_\tau-X_{\tau'}|$,  which is satisfied by the $X_\tau$ correlations that we find below. Since only the term $\phi_i(0,\tau)$ appears in the coupling, we can integrate the Bose fields at all other points $\phi_i(x\neq 0, \tau)$,  leading to the well studied $\sim|\omega|$ term \cite{giamarchi}, hence
 the total action becomes
\beq{06}
S&=&\half\int_\omega \{M\omega^2|X_\omega|^2+\sum_{i=1}^N \frac{2|\omega|}{\pi K}|\phi_\omega^i|^2\}\nonumber\\&-&g\int_\tau \sum_{i=1}^N \cos(X_\tau-2\phi_\tau^i)
\eeq
where $\phi_i(0,\tau)=\phi^i_\tau$ and its Fourier transform is $\phi^i_\omega$.
We now denote $B_\tau^i=2\phi^i_\tau-X_\tau$ as the fields entering in the non-linear term and
define $\tilde X_\omega$ in Fourier via
\beq{07}
X_\omega=\tilde X_\omega - \frac{1}{N_\omega}
\sum_{i=1}^N B_\omega^i \, , \quad N_\omega = N + 2 \pi MK |\omega|\label{6}
\eeq
where $N_\omega$ can be thought as an effective number of degrees of freedom.  It is then easy to see
that the action (\ref{e06}) can be rewritten as
a sum over two independent sectors, the field $\tilde X$ on one hand, and the $B_i$'s
on the other, as
\beq{08}
&&S=\half\int_\omega \{\frac{|\omega|}{2\pi K}N_\omega|\tilde X_\omega|^2+D_{i,j}^{-1}B_\omega^i B_\omega^{j*}\}
-g\sum_{i=1}^N \int_\tau \cos B_\tau^i  \nonumber\\
&&D_{i,j}^{-1}=
\frac{|\omega|}{2\pi K} (\delta_{i,j} - \frac{1}{N_\omega}) \, , \quad   D_{i,j}=\frac{1}{M\omega^2}+\frac{2\pi K}{|\omega|}\delta_{i,j}
\eeq
Hence one can first study the problem defined by the $B_i$ fields, and
in a second stage obtain the position of the particle $X_\tau$ from \eqref{6} as
the sum of two independent terms.  This decomposition immediately leads to two exact bounds,
first \cite{footnote}
\beq{09}
\langle |X_\omega|^2 \rangle \geq \langle |\tilde X_\omega|^2 \rangle = \frac{2 \pi K}{\omega N_\omega}
\eeq
where $\langle...\rangle$ denotes average over the action $S$. Furthermore
\beq{10}
\langle \cos X_\tau \rangle  \lesssim \langle \cos \tilde X_\tau \rangle\sim \left(\frac{4\pi^2KM}{N\beta}\right)^{K/N}\xrightarrow{\beta\to \infty} 0
\eeq
Hence the finite $N$ behavior of $\langle\cos X_\tau\rangle$ differs from the $N\to \infty$ case \cite{bh} where it can be finite and then serve as an order parameter.

To 0-th order in $g$, with $\langle...\rangle_0$ denoting an average with respect to $S_{g=0}$,
\beq{11}
\langle \cos B_\tau^i\rangle_0=\eexp{-\half\int_\omega \{ \frac{2\pi K}{|\omega|}+\frac{1}{M\omega^2}\}}=0
\eeq
which is {\it strongly irrelevant}
and cannot lead to an ordering of each individual $B^i_\tau$. Naively, one
could conclude from power counting that the coupling $g$ is washed away
by fluctuations, leading effectively to the $g=0$ theory.

However, this is not the case, as we have found: although strongly irrelevant, the terms $g \cos B_\tau^i$ generate
an effective coupling $g_2 \cos( B_\tau^i- B_\tau^j )$ between pairs of distinct fields. Indeed, the effective action evaluated to second order in perturbation theory in $g$ contains a term $\cos( B_\tau^i- B_{\tau'}^j )$ multiplied by
\beq{12}
 g^2\langle \eexp{iB_\tau^i-iB_{\tau'}^j}\rangle_0^{i\neq j}&=&g^2\eexp{-\frac{1}{2M}|\tau-\tau'|-\int_{\omega}\frac{2\pi K}{|\omega|}} \nonumber \\
& \rightarrow& \delta(\tau-\tau')M g^2\eexp{-\int_{\omega}\frac{2\pi K}{|\omega|}}
\eea
and integrated over times. We note that the finite mass is crucial to provide a short time cutoff $\sim M$.
The action involving the $B^i_\omega$ fields (denoted as $\bf B_\omega$) can thus be replaced by the effective action
\beq{13}
S_1=\half \int_\omega D_{i,j}^{-1}B_\omega^i B_\omega^{j*}- g_2 \Lambda \sum_{\VV}
\int_\tau \eexp{i\VV\cdot\BB_\tau}
\eeq
where $g_2\Lambda\sim Mg^2$, $g_2$ is a running dimensionless coupling, and $\Lambda$ a high frequency cutoff with initial value $\sim M$.  The vectors $\VV$ are N dimensional, have one entry of $+1$, one of $-1$ and all other entries are $0$, i.e. $\VV\cdot\BB_\tau=B^i_\tau-B^j_\tau$ with $i\neq j$. Hence $\VV$ form the primitive unit cell of an $N-1$ dimensional lattice that is perpendicular to the vector (1,1,...,1) on a simple cubic N dimensional lattice.
This type of model appears in various contexts, e.g. 
the quantum Brownian
motion in a periodic potential  \cite{yi1,yi2}. To 2nd order the RG flow equation is (see \cite{SM})
\beq{14}
  \Lambda \partial_\Lambda g_2 = (1- 2 K) g_2 + \alpha(N-2) g_2^2 + O(g_2^3)
\eeq
where $\alpha=O(1)>0$ is nonuniversal, depends on a smooth cutoff procedure.
Note that the TLL parameter $K$ is not renormalized and does not flow. From \eqref{e14}
there is clearly a critical line for $K > 1/2$ and $N>2$
\beq{15}
g_2^c= \frac{2(K-\half)}{\alpha(N-2)}
\eeq
such that for $g_2 < g_2^c$ the RG flow is towards the  Gaussian $g_2=0$ theory, while
for $g_2>g_2^c$, $g_2$ flows to strong coupling, signaling a phase where the relative
fields $B^i_\tau$  lock together, in a way that we
study below.

 The N=2 case has a single nonlinear term $\sim\cos(B^1_\tau-B^2_\tau)$, equivalent to the static impurity problem \cite{giamarchi,fateev,kane} and has a vertical phase boundary at $K=\half$ (Fig. 1).
 Going back to general $N$, we  now study the $B^i_\tau$ correlations by adding a source term $-\int_\omega |\omega| {\bf B}_\omega\cdot \AB_{-\omega}$ to the action (\ref{e13}) so that $\langle B^i_\omega B^j_{-\omega}\rangle=\frac{1}{Z_1\omega^2}
\frac{\delta^2Z_1}{\delta A^i_{-\omega}\delta A^j_\omega}|_{\AB=0}$, where \cite{footnote}
$Z_1=\int {\cal D}{\bf B} e^{-S_1}$ is the partition sum in presence of the source. Before studying the general correlations, we note \cite{SM} an exact sum rule of the effective model \eqref{e13} $\sum_i \langle B^i_\omega B^j_{-\omega}\rangle=N_\omega/(M\omega^2)$ for each $j$.

We proceed to study the strong coupling fixed point by a duality transformation. The process is well known for the $N=2$ case \cite{kane,giamarchi}, results are also stated for the quantum Brownian motion \cite{yi1,yi2}, yet the extension to $N>2$ of our case involves some subtleties. We perform first a change of variables so that the Gaussian part of $S_1$, Eq. (\ref{e13}), becomes diagonal,
\beq{17}
C^i_\omega=B^i_\omega-\alpha_\omega\bar B_\omega\,,\qquad \alpha_\omega=1 - \sqrt{1-N/N_\omega}
\eeq
where $\bar B_\omega=\sum_iB^i_\omega/N$ and $\bar C_\omega=\sum_iC^i_\omega/N=(1-\alpha_\omega)\bar B_\omega$. The action becomes
\beq{18}
S_1&=&\half \int_\omega \frac{|\omega|}{2\pi K}\CC_\omega\cdot\CC_{-\omega}- g_2\Lambda\sum_{\VV}\int_\tau \eexp{i \VV\cdot \CC_\tau}\nonumber\\
&-&\int_\omega |\omega|[\CC_\omega \cdot\AB_{-\omega}+\frac{\alpha_\omega}{1-\alpha_\omega}\bar C_\omega\sum_i A^i_{-\omega}]
\eeq

We consider next large $g_2$ where the trajectories of $\CC_\tau$ are dominated by instantons, i.e. a sequence of $n$ sharp jumps at consecutive times $\tau_1,\tau_2,...,\tau_\alpha,...,\tau_n$. The instantons shift $\CC_\tau$ between neighboring equivalent minima of the $g_2$ term by vectors chosen from a set $\RR_i$ such that
$\RR_i\cdot\VV_j=\delta_{i,j}$. Hence $\RR_i$ form the reciprocal lattice to $\VV_j$, each vector has one entry of $-1+1/N$ and all the rest are $1/N$, with norm $|\RR_i|^2=1-1/N$. The vectors $\RR_i$ are also orthogonal to (1,1,1,...), however they do not form a  primitive unit cell for $N>3$ and then their lattice symmetry differs from that of the $\VV_i$. E.g., for $N=3$ both $\VV_i,\RR_j$ form a 2D triangular lattice, however, for N=4 there are 12 vectors $\VV_i$ forming an fcc lattice while there are 8 vectors $\RR_i$ that form a bcc lattice.

Since $\RR_\alpha$ are perpendicular to (1,1,1,...) instanton trajectories do not describe the center of mass $\bar C_\omega$. Decoupling this center of mass is achieved by the shift $\tilde C^i_\omega=C^i_\omega-\bar C_\omega$, hence the Gaussian part in Eq. (\ref{e18}) decouples into $\CC_\omega\cdot\CC_{-\omega}=\tilde\CC_\omega\cdot\tilde\CC_{-\omega}+N |\bar C_\omega|^2$. The $\tilde \CC_\tau$ trajectory is described by
$\tilde \CC(\tau)=2\pi\sum_\alpha \RR_\alpha \theta(\tau-\tau_\alpha)$.
The coupling with the source can be written as $\int_\omega |\omega| {\bf\tilde C}^i_\omega \cdot {\bf A}^i_{-\omega}=i 2\pi \RR_\alpha \cdot \ab(\tau_\alpha)$ where $\ab_\omega=-\mbox{sign}\omega\AB_\omega$.
The weight of each instanton is defined as $\Lambda\Delta\sim\eexp{-S_{ins}}$ where \cite{kane,giamarchi}  $S_{ins}\sim \sqrt{g_2}$. In the strong coupling limit $\Delta=0$, instantons are absent and the correlations become
\beq{20}
\langle B^i_\omega B^j_{-\omega}\rangle&=&\frac{1}{(1-\alpha_\omega)^2}\langle |\bar C_\omega|^2\rangle=\frac{N_\omega}{NM\omega^2}
\eeq
so that all TLLs becomes equally coupled to each other.
If $\Delta>0$ the term $\half\int_\omega\frac{|\omega|}{2\pi K}\tilde\CC_\omega\cdot\tilde\CC_{-\omega}$ produces, after integration on $\omega$, logarithmic interactions between instantons \cite{SM} which correspond to the dual action,
\beq{21}
S_2=\half\int_\omega \frac{K|\omega|}{2\pi}|\bm\theta(\omega)|^2-\Lambda\Delta\sum_\RR\int_\tau \eexp{i\RR\cdot({\bm \theta}(\tau)+ 2\pi\ab(\tau))}\nonumber\\
\eeq

By shifting $\bm\theta\rightarrow\bm\theta(\tau)-2\pi\ab(\tau)$ and taking a 2nd derivative in $\AB_\tau$ we obtain a relation between the $\BB_\omega$ and $\bm\theta_\omega$ correlations
\beq{22}
\langle  B^i_\omega B^j_{-\omega}\rangle=\frac{N_\omega}{NM\omega^2}+\frac{2\pi K}{|\omega|} \delta_{i,j}
-K^2\langle \theta^i_\omega\theta^j_{-\omega}\rangle
\eeq

The dual form allows for deriving the RG equation, using $|\RR|^2=1-1/N$, to first order,
\beq{23}
\Lambda\partial_\Lambda\Delta=1-\frac{1}{K}(1-\frac{1}{N})\Delta
\eeq
Hence the phase transition at strong coupling terminates at $K_c=1-\frac{1}{N}$. The next order in RG for N=3 is $\sim\Delta^2$ (similarly to Eq.  (\ref{e14}) in the dual coupling $g_2$), while for $N\geq 4$ there are no $\Delta^2$ terms since $\RR\pm\RR'$ are all longer than $\RR$ and are therefore irrelevant at the transition. The next order is then $O(\Delta^3)$, hence the critical line at strong coupling is $\Delta_c\sim\sqrt{K_c-K}$ with an infinite slope at $K_c$, for $N\geq 4$. This is similar to the $N\rightarrow\infty$ case \cite{bh} where $K_c=1$ and $g_c^2\sim 1/(1-K)$. The various phase boundaries are illustrated in Fig. 1.

The $N=3$ case is self dual, i.e. we find a relation \cite{SM} between $\langle B^i_\omega B^j_{-\omega}\rangle_{K,g_2}$ and $\langle B^i_\omega B^j_{-\omega}\rangle_{K\rightarrow K/3,g_2\rightarrow\Delta}$. In particular at the self dual point  $K=1/\sqrt{3},g_2=\Delta$ on the critical line $\langle B^i_\omega B^j_{-\omega}\rangle$ is exactly
given by the average of its values for $g_2=0$ and for $\Delta=0$ (Fig. 1).

\hide{
The $N=2,3$ cases have a self dual lattice. For $N=2$ the phase boundary is vertical, hence this information is not so useful. For $N=3$ upon rescaling $\bm\theta\rightarrow (||\VV||/||\RR||)\bm\theta=\sqrt{3}\bm\theta$ yields $\langle \theta^i_\omega \theta^j_{-\omega}\rangle=3\langle C^i_\omega C^j_{-\omega}\rangle_{K\rightarrow K/3,g_2\rightarrow\Delta}$ and by using Eq. (\ref{e22}) we can derive a relation between $\langle B^i_\omega B^j_{-\omega}\rangle_{K,g_2}$ and $\langle B^i_\omega B^j_{-\omega}\rangle_{K\rightarrow K/3,g_2\rightarrow\Delta}$. In particular, at the self dual point $K=1/\sqrt{3},g_2=\Delta$ on the critical line we have an exact result \cite{SM}
\beq{24}
\langle  B^i_\omega B^j_{-\omega}\rangle=
\half\{\frac{N_\omega}{NM\omega^2}
+\frac{2\pi K}{|\omega|}\delta_{i,j}+\frac{1}{M\omega^2}\}
\eeq
i.e. precisely the average of the $g_2=0$ and $\Delta=0$ limits.}

We proceed now to identify the order parameter of our phase transition, i.e. the transconductance. The phenomenon of current in chain $i$ induced by a voltage on chain $j$ has been studied in the context of crossed nanotubes \cite{flensberg,nazarov,klesse,ponomarenko,komnik,gao}. In our case transconductance is a spontaneous order parameter and not a mechanical junction as for the nanotubes.
The usual experiment is a 2-probe type that for a single clean TLL yields \cite{tarucha,yacobi} a conductance $\frac{e^2}{h}$ 
determined by the normal leads
\cite{safi,maslov,ponomarenko2}. For our system of N TLLs in the decoupled phase obviously $G_{ij}=\frac{e^2}{h}\delta_{ij}$ while in the coupled phase the strong generated coupling $\cos[\phi_i(0,\tau)-\phi_j(0,\tau)]$ forces the currents $I_i=\dot\phi_i(0,\tau)$ to be equal, $I_i=I$ with total dissipation $NI^2\frac{h}{e^2}$. We propose then that, with normal leads on each TLL, the resistance measured by a voltage in one wire is the sum of all individual resistances, hence
 \beq{25}
G_{i,j}=\left\{\begin{array}{c}\frac{e^2}{h}\delta_{i,j}\qquad g_2=0\\
   \frac{e^2}{h}\frac{1}{N}\qquad \Delta=0\\
\end{array}\right.
\eeq
This implies that the transconductance exhibits a jump at the phase transition 
between these two values (Fig.1).

To substantiate this rationale, we consider first a "local conductance" for the response to a field applied on a length $L$ of a pure TLL \cite{giamarchi}. The response function away from the impurity involves \cite{SM} $\eexp{\pm|\omega_n|x/u}$, a constant in the DC limit. Hence $L\rightarrow 0$ can be taken, yielding
\beq{26}
G^{local}_{ij}(\omega)=\frac{-e^2}{\pi^2 \hbar}i(\omega+i\delta)\langle
\phi_i(\omega_n)\phi_j(-\omega_n)\rangle|_{i\omega_n\rightarrow \omega+i\delta}\nonumber
\eeq
In terms of the fields $B^i_\omega,\tilde X_\omega$ this becomes
\beq{27}
G^{local}_{i, j}(\omega)=\frac{e^2}{2\pi h}\omega[\langle B^i_\omega B^j_{-\omega}\rangle-\frac{1}{M\omega^2}]
\eeq
From the sum rule on the $B^i_\omega$ correlations we obtain the exact sum rule $\sum_iG^{local}_{i,j}=\frac{e^2}{h}K$. Using our results for the correlations, we obtain the DC local conductance at the fixed points and at the self dual point
\beq{28}
G^{local}_{i,j}=\left\{\begin{array}{c}\frac{e^2}{h}K\delta_{i,j}\qquad g_2=0\\
   \frac{e^2}{h}\frac{K}{N}\qquad \Delta=0\\
\half\frac{e^2}{h}K[\delta_{i,j}+\frac{1}{N}]\qquad \mbox{self-dual}\,\, (N=2,3)
\end{array}\right.
\eeq
Along the transition line the conductance varies continuously: the correction to $G^{local}_{i,j}$
is proportional to $1-N\delta_{i,j}$ with a positive prefactor $\sim (K-\half)^2$ near
$g_2=0$ and a negative one $\sim -(\frac{2}{3}-K)^2$ for $N=3$ and $\sim -(1-\frac{1}{N}-K)$
for $N \geq 4$ near $\Delta=0$ [\cite{SM} Eq. (49)]. The extension to the inhomogeneous case with normal leads is shown in \cite{SM}; following ideas of the N=1 case \cite{safi,maslov,ponomarenko2} results in replacing $K\rightarrow 1$, yielding Eq. \eqref{e25}.

We have also considered an $N=2$ case with
two coupled LLs, one with normal leads and the other a homogenous periodic TLL. We find \cite{SM} the conductance matrix $G_{ij}=\frac{e^2}{h}$, which also follows from our rationale since the TLL loop by itself has vanishing resistance. A similar problem was considered in \cite{chou} (see comparison in \cite{SM}).

We consider next the realization of our model by a superconducting (or BEC) grain. The Josephson coupling to s wave pairs in each TLL involves \cite{giamarchi} $\half g\eexp{iX_\tau-i\sqrt{2}\theta_{\rho,i}(0,\tau)}$ where $X_\tau$ is now the superconducting phase of the grain and $\theta_{\rho,i}$ is the dual phase to $\phi_{\rho,i}$ in the charge sector of chain $i$. The action in terms of $\theta_{\rho,i}$ has the same form as in Eqs. (\ref{e03},\ref{e04}) with $K\rightarrow 1/(2K_\rho)$ \cite{giamarchi} and $\frac{1}{2M}$ being the charging energy $E_c$ of the grain; hence the phase diagram is also given by Fig. 1 with the axis being $1/(2K_\rho)$. Thus, for $N=2$ the phase boundary is at $K_\rho=1$  and the strong coupled phase appears even for weakly attractive coupling $K_\rho>1$. We note that the data \cite{bouchiat} on a single wall carbon nanotube, expected to have $N=2$, shows with superconducting leads a surprisingly high supercurrent. If one of the leads contains a grain with not too small charging energy then our strong coupling phase, implying a strong Josephson coupling, would account for the data. For $N>2$, possible for nanotube ropes \cite{bouchiat}, the phase boundary interpolates between $K_\rho=1$ and $K_\rho=\frac{N}{2(N-1)}$, allowing for a relevant Josephson coupling even in a range of repulsive interactions. The transconductance of this case needs a separate derivation \cite{SM}, yet the result is the same as Eq. \eqref{e28} except $K\rightarrow 2K_\rho$ and a prefactor 2 for this spinfull case.

Finally,
from the decomposition (\ref{e07}) and the sum rule for the $B^i_\omega$ correlations within the effective model Eq. (\ref{e13}), we obtain the fluctuations of $X_\tau$ as
$\langle |X_\omega|^2\rangle=\frac{1}{M\omega^2}$
i.e. they are not affected by the phase transition. Therefore $\langle (X_\tau-X_{\tau'})^2\rangle\sim |\tau-\tau'|$ justifies our assumption in deriving the action (\ref{e06}), i.e. that $|X_\tau-X_{\tau'}|\ll |\tau-\tau'|$.

We note that a finite impurity mass is essential for the derivation of our effective action (\ref{e08}), although it does not appear explicitly in the phase diagram.
 As seen from Eqs. (\ref{e11},\ref{e12}) $1/M$ provides an upper limit on frequencies which implies an upper bound on temperature in a possible experiment, $T^*\approx \frac{1}{M}= (2\pi\rho_0)^2/M_0$.
 For Cs atoms \cite{meinert} and TLL density $\pi \rho_0=4.5 \mu m^{-1}$ we find $T^*=10^{-7}K$; increasing the TLL density or reducing $M_0$ increase the range of $T<T^*$. For the realization with a superconducting grain $T^*\approx E_c$ where $E_c\approx 1-10K$ \cite{marcus} is achievable in such devices.
  For $M\rightarrow\infty$ the problem reduces to a static impurity \cite{kane} with a phase transition at $K=1$ that separates a conducting phase from an insulating phase and no transconductance.

To realize a cold-atom experiment, one could consider an optical trap array of parallel tubes for the TLLs (as in Ref. \onlinecite{meinert}) and impurity atoms residing at the centers of the array's unit cells. The latter is possible by choosing the trapping frequency to be simultaneously red detuned for the TLL atoms and blue detuned for the impurity atoms, or vise versa, producing opposite sign couplings to the laser intensity; this arrangement was actually utilized \cite{lercher} for Rb-Cs mixtures. To produce a reasonable impurity-TLL coupling we propose two routes. First, use atoms with a dipole moment (e.g. as in Er or Dy \cite{ilzhofer}). The long range dipole-dipole interaction provides the impurity-TLL interaction. A second route, with short range interactions,  is to produce a cage type trap for the impurities that is shallow inside and allows a reasonable overlap with the TLL atoms, yet has barriers to keep the impurity in a given unit cell. The impurity-TLL interaction could then be enhanced by a Feshbach resonance as for the K-Rb case \cite{catani}.

In conclusion, we have found that a dynamic boundary, such as a mobile impurity or a superconducting (or BEC) grain, coupled to $N$ identical Luttinger liquids induces a phase transition for all $N\geq 2$. The order parameter is the conductance matrix, in particular a large transconductance appears in the strong coupling phase.  In the superconducting grain case the strong coupling phase is also identified by a strong Josephson coupling, relevant to a number of active experimental setups \cite{bouchiat,marcus,kouwenhoven}.

\acknowledgements
BH thanks Y. Meir, E. Sela, I. Lerner, R. Folman and G. Zar\'and for stimulating discussions and Netanel Mirilashvili for his partaking in the $1/N$ expansion. BH also gratefully acknowledges funding by the German DFG through the DIP programme [FO703/2-1]. PLD acknowledges support from ANR grant ANR-17-CE30-0027-01 RaMaTraF. This work was supported in part by the Swiss National Science Foundation under Division II.

\begin{widetext}

\bigskip

\bigskip

\begin{large}
\begin{center}

SUPPLEMENTARY MATERIAL

\end{center}
\end{large}

\bigskip

We present here details of the calculations, as
described in the main text of the Letter. The topics are RG equations, correlation functions, duality, conductance -- either local or with normal leads in various geometries, and a complementary $1/N$ expansion.

\section{I. Renormalization Group}

We derive here the RG equation (14) of the main text. The 1st order RG is identified directly from Eq. (12) of the main text
\beq{101}
&&g_2^R\Lambda'=g_2\Lambda \eexp{-\int_{\Lambda'}^\Lambda\frac{2\pi K}{|\omega|}\frac{d\omega}{\pi}}=g_2\Lambda(1-2K\frac{d\Lambda}{\Lambda})\nonumber\\
&&\Rightarrow g^R_2=g_2(1+(1-2K)\frac{d\Lambda}{\Lambda}))
\eeq
where $d\Lambda=\Lambda-\Lambda'$ is infinitesimal. To 2nd order the partition function is (subtracting the counterterm, i.e. the 2nd order correction from exponentiating Eq. \eqref{e101})
\beq{102}
Z^{(2)}&=&\half(2g_2\Lambda)^2\int_{\tau,\tau'}\langle\sum_{i\neq j}\cos(B^i_\tau-B^j_\tau)\sum_{k\neq l}\cos(B^k_{\tau'}-B^l_{\tau'})
\rangle_{\Lambda'}^{\Lambda}\nonumber\\&-&\half(2g_2\Lambda)^2\eexp{-4K\int_{\Lambda'}^\Lambda d\omega/\omega}\sum_{i\neq j}\cos(B^i_\tau-B^j_\tau)\sum_{k\neq l}\cos(B^k_{\tau'}-B^l_{\tau'})
\eeq
where $\langle ...\rangle^\Lambda_{\Lambda'}$ means average over frequencies in the range $[\Lambda',\,\Lambda].$
Collect first the indices such that only one pair are equal, e.g. $l=j,\,k\neq i$. We need the average $\langle...\rangle$ defined here in the frequency range $\Lambda',\,\Lambda$
\beq{103}
\half(g_2\Lambda)^2\int_{\tau,\tau'}\sum_\pm \langle \eexp{i(B^i_\tau-B^j_\tau)\pm i(B^j_{\tau'}-B^k_{\tau'})}+h.c.\rangle
\eeq
The average for the upper sign gives an exponent of
\beq{104}
&&\half\langle(B^i_\tau-B^j_\tau+ B^j_{\tau'}-B^k_{\tau'})^2\rangle=\int^\Lambda_{\Lambda'}\frac{d\omega}{2\pi}[2D_{ii}-2D_{i\neq j}+D_{i\neq j}\eexp{i\omega(\tau-\tau')}-D_{i\neq k}\eexp{i\omega(\tau-\tau')}\nonumber\\&-&D_{jj}\eexp{i\omega(\tau-\tau')}+D_{j\neq k}\eexp{i\omega(\tau-\tau')}+c.c]=\int^\Lambda_{\Lambda'}\frac{d\omega}{2\pi}[4\frac{2\pi K}{\omega}-2\frac{2\pi K}{\omega}\cos\omega(\tau-\tau')]
\eeq
Together with the counterterm in this group
\beq{105}
Z^{(2a)}=(g_2\Lambda)^2\int_{\tau,\tau'}\cos(B^i_\tau-B^j_\tau+ B^j_{\tau'}-B^k_{\tau'})\eexp{-4K\int_{\Lambda'}^\Lambda\frac{d\omega}{\omega}}
[\eexp{2K\int_{\Lambda'}^\Lambda\frac{\cos\omega(\tau-\tau')}{\omega}d\omega}-1]
\eeq
The [...] factor is short range (at least after choosing a smooth cutoff, as done in the similar 2D sine-Gordon \cite{kogut,ohta,knops}) and is positive. We then replace $\tau'\rightarrow \tau$ in the integrand and the $\tau'$ integral becomes $\alpha\frac{d\Lambda}{\Lambda^2}$ where $\alpha=O(1)>0$ is dimensionless and depends on the cutoff procedure. Hence a term $\sum_{i\neq k}\cos[B^i_\tau-B^k_\tau]$ is generated, and since there are $N-2$ choices for $j$, the 2nd order RG yields Eq. (14) of the main text.

The lower sign in \eqref{e103} generates a term
$\tilde g\Lambda\int_\tau \cos[B^i_\tau+B^k_\tau-2B^j_\tau]$ whose 1st order RG is
\beq{106}
\tilde g\Lambda\langle\int_\tau \cos[B^i_\tau+B^k_\tau-2B^j_\tau]\rangle^\Lambda_{\Lambda'}=\eexp{-\half\int_\omega(6D_{ii}-6D_{i\neq j})}
=\eexp{-6\int^\Lambda_{\Lambda'}\frac{d\omega}{2\pi}\frac{2\pi K}{\omega}}
\eeq
The RG equation for $\tilde g$ to 1st order is
\beq{107}
\tilde g^R=\tilde g[1+(1-6K)\frac{d\Lambda}{\Lambda}]
\eeq
Hence $\tilde g$ is relevant only at $K<1/6$, irrelevant for the phase transition of $g_2$.

Consider next the case of two pairs of equal indices, i.e. $\cos(B^i_\tau-B^j_\tau+B^j_{\tau'}-B^i_{\tau'})$. As above, there is a short range kernel that causes the argument of the $\cos$ to vanish; a next order expansion in $\tau-\tau'$ can yield $[\partial_\tau(B^i-B^j)]^2$ which is $\sim \omega^2$, hence negligible in comparison with $D^{-1}\sim |\omega|$. Finally, if all indices are distinct we need
\beq{108}
\langle(B^i_\tau-B^j_\tau+B^k_{\tau'}-B^l_{\tau'})^2\rangle=&&\int_{\Lambda'}^\Lambda\frac{d\omega}{2\pi}[2D_{ii}-2D_{i\neq j}
+D_{i\neq k}\eexp{i\omega(\tau-\tau')}-D_{i\neq l}\eexp{i\omega(\tau-\tau')}-D_{j\neq k}\eexp{i\omega(\tau-\tau')}\nonumber\\
&&+D_{j\neq l}\eexp{i\omega(\tau-\tau')}+c.c]=4K\int_{\Lambda'}^\Lambda\frac{d\omega}{\omega}
\eeq
which is exactly canceled by the counterterm. Hence the only needed renormalization is Eq. \eqref{e105}, leading to Eq. (14) of the main text.

The case $N=2$ can be discussed separately. One can define
$B^{\pm}_\tau = B^1_\tau \pm B^2_\tau$. The sum
$B^+_\tau$ is Gaussian and decouples, while the action for the difference field becomes
\beq{109}
S_{N=2}^-=\half\int_\omega \frac{|\omega|}{4\pi K}|B^-_\omega|^2 -g_2\int_\tau\cos(B^-_\tau)
\eeq
This is the well studied static impurity problem \cite{giamarchi,fateev,kane},
that has a phase transition at $K=\half$, independent of $g_2$, i.e. with a vertical
phase boundary.

\section{II. Correlation functions}

{\it Notational remark:} In the main text we use the shorthand notation $\langle B^i_\omega B^j_{-\omega}\rangle$ to denote $\lim_{T\rightarrow 0} \{T\langle B^i_{\omega_n} B^j_{-\omega_n}\rangle\}$ (with $\omega_n$  Matsubara frequencies)
whereas the complete $T=0$ correlation in continuum frequencies (denoted with a prime) is $\langle B^i_\omega B^j_{\omega'}\rangle'=2 \pi \delta(\omega+\omega')\langle B^i_\omega B^j_{-\omega}\rangle$. In the Supplementary Material we use the form with Matsubara frequencies, though for brevity we denote
$\omega_n\rightarrow \omega$.

In this section we introduce source terms, evaluate the $B^i_\omega$ correlations and derive the sum rule mentionned below Eq. (16) in the main text. Consider the effective action Eq. 13 in the main text and add a source ${\bf A}_\tau$ (related to a vector potential)
\beq{110}
&&S_1=\half \int_\omega D_{i,j}^{-1}B_\omega^i B_\omega^{j*}-  g_2\Lambda\sum_{\VV}\int_\tau \eexp{i \VV\cdot \BB_\tau}-\int_\omega |\omega|\BB_\omega \cdot\AB_{-\omega}\nonumber\\
&&=\half \int_\omega \frac{|\omega|}{2\pi K}\{\BB_\omega\cdot\BB_{-\omega}-\frac{1}{N_\omega}|\sum_i B^i_\omega|^2\} -  g_2\Lambda\sum_{\VV}\int_\tau \eexp{i \VV\cdot \BB_\tau}-\int_\omega |\omega|\BB_\omega \cdot\AB_{-\omega}
\eeq
The set of vectors $\VV$ are all N-component vectors with
one entry $+1$, one entry $-1$ and all other entries $0$. It is easy
to see that they are all minimal norm vectors of the Bravais lattice
generated by $N-1$ basis vectors, labeled as $\VV_\nu$, $\nu=1,..N-1$
and which can be chosen as:
\beq{111}
\VV_1= (1,-1,0,0,...) \quad , \quad
\VV_2=(1,0,-1,0,...) \quad , \quad
\VV_3=(1,0,0,-1,...)\quad , \quad  ...
\eeq
Their coordinates are denoted $(\VV_\nu)_i \equiv V_\nu^i$ with
$V_\nu^i = \delta_{1,i}-\delta_{\nu+1,i}$, $i=1,..,N$.
These vectors are all perpendicular to (1,1,1,1,...), hence the set $\VV_\nu$, $\nu=1,..N-1$, is a primitive unit cell of an N-1 dimensional space, a "hypertriangular" lattice \cite{yi2}, i.e. all lattice points of a simple cubic N dimensional lattice that are perpendicular to (1,1,1,1,....). For N=3 it is a 2D triangular lattice, for N=4 these form tetraheders which are the primitive cell for an fcc lattice.

The partition sum of $S_1$ generates the correlation function needed later, using $\int_\omega=T\sum_{\omega_n}$,
\beq{112}
\langle  B^i_\omega B^j_{-\omega}\rangle=\frac{1}{Z_1}\frac{1}{\omega^2 T^2}\frac{\delta^2 Z_1}{\delta  A^i_{-\omega}\delta  A^j_{\omega}}|_{A=0}
\eeq
All correlations denoted $\langle...\rangle$ here and below imply $\AB=0$.
An equivalent form of this correlation is obtained by writing the Gaussian terms as
$\frac{|\omega|}{4\pi K}|(\BB-2\pi K\AB)_{\omega}|^2-\pi K|\omega||\AB_\omega|^2$
and shifting $\BB_\omega\rightarrow\BB_\omega+2\pi K\AB_\omega$,
\beq{113}
S_1&=&\half\int_\omega\frac{|\omega|}{2\pi K} \{\BB_\omega\cdot  \BB_{-\omega}-\frac{1}{N_\omega}|\sum_iB^i_\omega+2\pi K A^i_\omega|^2\} -\pi K\int_\omega|\omega||\AB_\omega|^2\nonumber\\
&-& g_2\Lambda\sum_{\VV}\int_\tau \eexp{i \VV\cdot (\BB_\tau+2\pi K A_\tau)}
\eeq
The partition sum $Z_1=\int {\cal D} \BB \eexp{-S_1}$ involves a path integral $\int {\cal D} \BB$, i.e. integration on all periodic functions $\BB_\tau$ (with period $\beta$, where we study mostly the limit $\beta \to +\infty$). Therefore, from \eqref{e112}
\beq{114}
&& \langle  B^i_\omega B^j_{-\omega}\rangle=\frac{1}{Z_1 T \omega^2}\frac{\delta}{\delta A_{\omega}^j}
\int_\BB \{\frac{|\omega|}{N_\omega}\sum_k(B_{\omega}^k+2\pi KA_{\omega}^k)+2\pi K|\omega|A_{\omega}^i
\nonumber\\&&+ g_2\Lambda \sum_\VV iV^i 2\pi K\int_\tau\eexp{i\omega\tau}\eexp{i\VV\cdot(\BB_\tau+2\pi K\AB_\tau)}\}\eexp{-S_1}|_{\AB=0}=
\frac{2\pi K}{T |\omega|}(\frac{1}{N_\omega}+\delta_{i,j})\nonumber\\&&+\frac{1}{N_\omega^2}\sum_{k,l}\langle B_{\omega}^k B_{-\omega}^l \rangle
-( g_2\Lambda)^2\frac{(2\pi K)^2}{\omega^2}\sum_{\VV,\VV'}
V^iV'^j\int_{\tau,\tau'}\eexp{i\omega(\tau-\tau')}\langle \eexp{i\VV\cdot \BB_\tau+i\VV'\cdot\BB_{\tau'}}\rangle\nonumber\\
&& -g_2\Lambda  \frac{(2\pi K)^2}{\omega^2}  \sum_{\VV}V^iV^j\int_\tau \langle\eexp{i\VV\cdot\BB_\tau}\rangle
\eeq
We use here $\int_\tau\eexp{-i\omega\tau}\langle \eexp{i\VV\cdot\BB_\tau}\sum_kB_{-\omega}^k\rangle=0$ which can be shown by using the variables $\tilde B_i(\tau)=B_i(\tau)-\bar B(\tau)$ where $\bar B_\tau=\frac{1}{N}\sum_iB_\tau^i$; the Gaussian part of the action becomes
\beq{115}
&&S_0=\half\int_\omega\sum_{i,j}\frac{|\omega|}{2\pi K}(\delta_{i,j} - \frac{1}{N_\omega})(\tilde B_i+\bar B)_\omega(\tilde B_j+\bar B)_{-\omega}=
\half\int_\omega\frac{|\omega|}{2\pi K} \tilde\BB_\omega\cdot \tilde \BB_{-\omega}+\half\int_\omega\frac{NM\omega^2}{N_\omega}|\bar B_\omega|^2\nonumber\\
\eeq
Since $\eexp{i\VV\cdot\BB_\tau}=\eexp{i\VV\cdot\tilde\BB_\tau}$ is independent of $\bar B_\tau$ we have $\int_\tau\eexp{-i\omega\tau}\langle \eexp{i\VV\cdot\BB_\tau}\sum_kB_{-\omega}^k\rangle\sim\langle \bar B_{-\omega}\rangle=0$.
Since $\sum_iV^i=0$ we obtain
 from \eqref{e114} an exact sum rule
\beq{116}
\sum_i T \langle  B^i_\omega B^j_{-\omega}\rangle=\frac{2\pi K}{|\omega|}(\frac{N}{N_\omega}+1)/(1-\frac{N^2}{N_\omega^2})=\frac{N_\omega}{M\omega^2}\,,\qquad
T \langle |\bar B_\omega|^2\rangle=\frac{N_\omega}{NM\omega^2}
\eeq
 Hence the $g_2$ independent terms of $T \langle  B^i_\omega B^j_{-\omega}\rangle$ are
\beq{117}
\frac{2\pi K}{|\omega|}(\frac{1}{N_\omega}+\delta_{i,j})+\frac{N}{N_\omega M\omega^2}=\frac{2\pi K}{|\omega|}\delta_{i,j}+\frac{1}{M\omega^2}=D_{i,j}
\eeq
which is consistent with the $g_2=0$ result. Hence finally we obtain the exact formula (for the effective model $S_1$)
\beq{118}
\langle  B^i_\omega B^j_{-\omega}\rangle=&&\frac{D_{i,j}}{T}-( g_2\Lambda)^2\frac{(2\pi K)^2}{\omega^2}\sum_{\VV,\VV'}
V^iV'^j\int_{\tau,\tau'}\eexp{i\omega(\tau-\tau')}\langle \eexp{i\VV\cdot \BB_\tau+i\VV'\cdot\BB_{\tau'}}\rangle\nonumber\\
&& -g_2\Lambda  \frac{(2\pi K)^2}{\omega^2} \sum_{\VV}V^iV^j\int_\tau \langle\eexp{i\VV\cdot\BB_\tau}\rangle
\eeq

We now use the latter form for a perturbative expansion of the correlation function in $g_2$. Consider first the last term of \eqref{e118}, to lowest order it has
\beq{119}
\Lambda\int_\tau \langle \eexp{i\VV\cdot\BB_\tau}\rangle_0=\Lambda\int_\tau\eexp{-\half\sum_{kl}V^kV^l\langle B^k_\tau B^l_\tau\rangle_0}=\Lambda\int_\tau\eexp{-\half\sum_k(V^k)^2\int\frac{2\pi K}{|\omega|}\frac{d\omega}{2\pi}}=\left(\frac{T}{\Lambda}\right)^{2K-1}{\overset{K>\half}
{\longrightarrow}}\, 0
\eeq
where $\langle ...\rangle_0$ is an average w.r.t. $S_1(g=0,\AB=0)$. We use temperature $T$ as a lower bound on the $\omega$ integral and $1/T$ is an upper bound on the $\tau$ integral. We note also that the factor $\frac{1}{M\omega^2}$ common to all $D_{kl}$ is canceled since $\sum_k V^k=0$ and that the perturbation expansion is valid only for $K>\half$.

The next order of the last term of \eqref{e118} combines with the 2nd term so that
 \[ T\langle  B^i_\omega B^j_{-\omega}\rangle=D_{i,j}-( g_2\Lambda)^2\frac{(2\pi K)^2 T}{\omega^2}\sum_{\VV,\VV'}
\int_{\tau,\tau'}[V^iV'^j\eexp{i\omega(\tau-\tau')}+V^iV^j]\langle \eexp{i\VV\cdot \BB_\tau+i\VV'\cdot\BB_{\tau'}}\rangle_0 +O(g_2^4)  \]
The average $\langle (\VV\cdot \BB_\tau+i\VV'\cdot\BB_{\tau'})^2\rangle_0$ involves $D_{ij}$ whose common term $1/M\omega^2$ cancels since $\sum_k V^k=0$. Thus only diagonal $D_{ii}$ survive with
$\frac{2\pi K}{|\omega|}$. Hence
\[\eexp{-\half \langle (\VV\cdot \BB_\tau+i\VV'\cdot\BB_{\tau'})^2\rangle_0}=
\eexp{-\half\sum_k[(V^k)^2+(V'^k)^2+2V^kV'^k\cos\omega(\tau-\tau')]\int_\omega\frac{2\pi K}{|\omega|}}\]
To cancel the infrared divergence, i.e. avoid the vanishing terms as in \eqref{e119},  only terms with $\VV=-\VV'$ survive,
\[
\langle\eexp{i\VV\cdot (\BB_\tau-\BB_{\tau'})}\rangle_0=\eexp{-\half \VV^2\int_{1/|\tau-\tau'|}^\Lambda\frac{4\pi K}{\omega}\frac{d\omega}{\pi}}=(\Lambda|\tau-\tau'|)^{-4K}\qquad |\tau-\tau'|>1/\Lambda
\]
Hence to order $O(g_2^2)$ we finally obtain,
\beq{120}
&&T \langle  B^i_\omega B^j_{-\omega}\rangle=D_{i,j}- ( g_2\Lambda)^2\frac{(2\pi K)^2}{\omega^2}\sum_\VV V^iV^j\int_{|\tau|>1/\Lambda}(1-\eexp{i\omega\tau})(\Lambda|\tau|)^{-4K}\nonumber\\
&& =D_{i,j}- ( g_2\Lambda)^2\frac{(2\pi K)^2}{\omega^2\Lambda}\sum_\VV V^iV^j\frac{2}{4K-1}(\frac{\omega}{\Lambda})^{4K-1}
\eeq

We can now use the previous RG study to express this result in terms of the renormalized
coupling $g_2^R$. For this purpose we consider $\omega$ as the new cutoff $\Lambda'=\Lambda-d\Lambda$, hence $(\frac{\omega}{\Lambda})^{4K-2}\rightarrow 1-(4K-2)\frac{d\Lambda}{\Lambda}$. This identifies $g_2^R$ to 1st order, using Eq. \eqref{e101}.
Therefore near $K=\half$,  we obtain to lowest order in $g_2^R$,
\beq{121}
T \langle  B^i_\omega B^j_{-\omega}\rangle=D_{i,j}+(g_2^R)^2\frac{4\pi^2}{|\omega|}(1-N\delta_{i.j})
\eeq
 where $\sum_\VV V^iV^j=
\sum_{a,b=1}^N (\delta_{ia}-\delta_{ib}) (\delta_{ja}-\delta_{jb})=
2(N\delta_{i,j}-1)$. We can apply this along the critical line Eq. (15) in the main text. One sees that the critical
correlation depends continuously on $g_2^R$.
For $N >2$ it yields a $K$-dependent correction,
$\sim(K-\half)^2 (1-N\delta_{i.j})/|\omega|$ to the critical correlation function,
$T \langle  B^i_\omega B^j_{-\omega}\rangle$, originating from the $g_2^R$ term.
\\

\section{III. Duality}

We derive here the duality between weak and strong coupling which is used in the main text.
We follow ideas stated by Yi and Kane \cite{yi1,yi2}. We also derive explicitly the basis vectors, and the correlation functions given in the text.

In the following we introduce the basis of the reciprocal lattice, i.e. $N-1$ vectors $\RR_\nu$ that satisfy $\RR_\nu\cdot\VV_{\nu'}=\delta_{\nu,\nu'}$
for $\nu,\nu'=1,..N-1$. Define a matrix $G_{i,\nu}=(\VV_\nu)_i$, i.e. the $i$-th component of the vector $\VV_\nu$. This has N rows $(i=1,2,...,N)$ and N-1 columns $(\nu=1,2,..., N-1)$. Similarly, define a matrix $R_{i,\nu}=(\RR_\nu)_i$. By definition
\beq{122}
&&\sum_i R_{i,\nu}G_{i,\nu'}=\delta_{\nu,\nu'}\,,\qquad \Rightarrow R^TG=1\qquad\\
&&\mbox{A solution is:}\qquad R^T=(G^TG)^{-1}G^T
\qquad \mbox{since} \qquad R^TG=(G^TG)^{-1}G^TG=1
\eeq
The norms are $|\RR_\nu|^2=\sum_i R_{i,\nu}^2=(R^TR)_{\nu,\nu}$,  so their sum satisfies
$\sum_i|\RR_i|^2 =TrRR^T=Tr G(G^TG)^{-1}(G^TG)^{-1}G^T=Tr (G^TG)^{-1}$. More explicitly, for N=4
\beq{123}
&&G^TG=\left( {\begin{array}{cccc} 1&-1&0&0\\1&0&-1&0\\1&0&0&-1\end{array}}\right)
\left( {\begin{array}{ccc} 1&1&1\\-1&0&0\\0&-1&0\\0&0&-1\end{array}}\right)=1+\delta_{i,j} \qquad (G^TG)^{-1}=\delta_{i,j}-\frac{1}{N}\nonumber\\
&&R=G(G^TG)^{-1}=\left( {\begin{array}{ccc} 1&1&1\\-1&0&0\\0&-1&0\\0&0&-1\end{array}}\right)
\left( {\begin{array}{ccc} 1-\frac{1}{N} & -\frac{1}{N} & -\frac{1}{N} \\
- \frac{1}{N} & 1-\frac{1}{N} & -\frac{1}{N} \\ -\frac{1}{N} & -\frac{1}{N} & 1-\frac{1}{N}\end{array}}\right)=
\left({\begin{array}{cccc} \frac{1}{N} & \frac{1}{N} & \frac{1}{N} \\
-1+\frac{1}{N} & \frac{1}{N} & \frac{1}{N}  \\
\frac{1}{N} & -1+\frac{1}{N} & \frac{1}{N}  \\
\frac{1}{N} & \frac{1}{N} & -1+\frac{1}{N} \end{array}}\right)\nonumber
\eeq
The following is valid for all $N$: $G^TG=1+\delta_{i,j},\, (G^TG)^{-1}=\delta_{i,j}-\frac{1}{N},\,
R_{i,\nu}=\frac{1}{N}-\delta_{i,\nu+1}$ and
the norm of each $\RR_\nu$ is $|\RR_\nu|^2=1-\frac{1}{N}$. The $\RR_\nu$ are also orthogonal to (1,1,1,...), however they are not a primitive unit cell for $N>3$ and therefore their lattice symmetry differs from that of the $\VV_\nu$
(see discussion in the text).
Note that for $N\rightarrow \infty$ it is an N-1 dimensional cubic lattice.
Finally note, as mentionned in the text, that in this reciprocal lattice there are $2N$ vectors $\RR$ of minimal norm $|\RR_\nu|^2=1-\frac{1}{N}$.

The duality relates $S_1$ to an action whose Gaussian part is diagonal. To apply duality we must change first the $B^i_\tau$ variables to $C^i_\tau$ such that $S_0$ is diagonal $\sim \mathbbm{1}$, with $\bar B_\omega=\sum_iB^i_\omega/N$,
\beq{124}
&&C^i_\omega=B^i_\omega-\alpha_\omega \bar B_\omega,\qquad \bar C_\omega=\sum_iC^i_\omega/N=(1-\alpha_\omega)\bar B_\omega \nonumber\\
&& \sum_{i,j}(\delta_{i,j}-\frac{1}{N_\omega})B_\omega^iB_{-\omega}^j=\sum_i(C_\omega^i+\alpha_\omega \bar B_\omega)(C_{-\omega}^i+\alpha_\omega\bar B_{-\omega})-
\frac{N^2}{N_\omega}|\bar B_\omega|^2\nonumber\\ &&=\sum_iC^i_\omega C^i_{-\omega}+N|\bar B_\omega|^2[\alpha_\omega^2+2\alpha_\omega(1-\alpha_\omega)-\frac{N}{N_\omega}]=\sum_iC^i_\omega C^i_{-\omega}\nonumber\\
&& \Rightarrow  \alpha_\omega=1 - \sqrt{1-\frac{N}{N_\omega}}
\eeq
Note that Det$[\frac{\partial C^i_\omega}{\partial B^j_\omega}]=$Det$[\delta_{i,j}-\frac{\alpha_\omega}{N}]=1-\alpha_\omega$, 
hence this transformation is proper if $\alpha_\omega\neq 1$, i.e. $\omega\neq 0$; therefore $\omega=0$ is used only as a limit. Using $\langle B^i_\omega\bar B_{-\omega}\rangle=\langle |\bar B_\omega|^2\rangle=\frac{N_\omega}{TNM\omega^2}$ it is useful to note
\beq{125}
\langle C^i_\omega C^j_{-\omega}\rangle=\langle B^i_\omega B^j_{-\omega}\rangle+(\alpha_\omega^2-2\alpha_\omega)\langle |\bar B_\omega|^2\rangle=
\langle B^i_\omega B^j_{-\omega}\rangle-\frac{1}{TM\omega^2}
\eeq

Using now $C^i_\omega=B^i_\omega-\frac{\alpha_\omega}{1-\alpha_\omega}\bar C_\omega$ we obtain the source term in terms of $C^i_\omega$ and the action becomes
\beq{126}
S_1=\half \int_\omega \frac{|\omega|}{2\pi K}\CC_\omega\cdot\CC_{-\omega}-  g_2\Lambda\sum_{\VV}\int_\tau \eexp{i \VV\cdot \CC_\tau}-\int_\omega |\omega|[\CC_\omega \cdot\AB_{-\omega}+\frac{\alpha_\omega}{1-\alpha_\omega}\bar C_\omega\sum_i A^i_{-\omega}]
\eeq

 Consider now large $g_2$ where the trajectories of $\CC_\tau$ are dominated by instantons, i.e. a sequence of n sharp jumps at consecutive times $\tau_1, \tau_2,..., \tau_\alpha,..., \tau_n$ that shift $\CC_\tau$ between neighboring equivalent minima by a vector $2\pi\RR_\alpha$, each chosen from the set of equivalent minimal length vectors  (here the index $\alpha$ labels the instanton and should not be confused with $\nu$ which labels the basis vectors).
Eventually there is a summation on all $n$ such that periodic boundary conditions (in imaginary time $\tau$) are maintained, i.e. $\sum_{\alpha}\RR_\alpha=0$, in particular $n$ is even.
Since all $\RR_\alpha$ are perpendicular to (1,1,1,...) the instanton trajectory describes only $\tilde C^i(\tau)=C^i_\omega-\bar C_\omega$. The action can be written as the sum of two independent parts,
\beq{127}
S_1=&&\half \int_\omega \frac{|\omega|}{2\pi K}\tilde\CC_\omega\cdot\tilde\CC_{-\omega}-  g_2\Lambda\sum_{\VV}\int_\tau \eexp{i \VV\cdot \tilde\CC_\tau}-\int_\omega |\omega|\tilde\CC_\omega \cdot\AB_{-\omega}\nonumber\\&&
+\half N\int_\omega \frac{|\omega|}{2\pi K}|\bar C_\omega|^2-\frac{|\omega|}{1-\alpha_\omega}\int_\omega \bar C_\omega\sum_i A^i_{-\omega}
\eeq
Note that evaluating directly the $\tilde C^i_\omega$ part is cumbersome since these are dependent variables, yet this form is convenient for the instanton description.
\beq{128}
&& \tilde \CC(\tau)=2\pi\sum_\alpha \RR_\alpha \theta(\tau-\tau_\alpha)=i2\pi\sum_\alpha\RR_\alpha\int_\omega \frac{\eexp{-i\omega(\tau-\tau_\alpha)}}{\omega+i\epsilon}\nonumber\\
&&\tilde \CC(\omega)=i2\pi\sum_\alpha\RR_\alpha\frac{\eexp{i\omega\tau_\alpha}}{\omega+i\epsilon}
\eeq
where $\epsilon=+0$. The source term with the $\tilde\CC_\omega$ part is
\beq{129}
&&\int_\omega |\omega| \tilde C^i_\omega A^i_{-\omega}= i2\pi \sum_\alpha R_\alpha^i\int_\omega \frac{|\omega|\eexp{i\omega\tau_\alpha}}{\omega+i\epsilon}A^i_{-\omega}\equiv i 2\pi \RR_\alpha \cdot \ab(\tau_\alpha)\nonumber\\
&&\ab_\omega=\frac{|\omega|}{-\omega+i\epsilon}\AB_{\omega}\,,\qquad \ab(\tau)=\int_\omega\frac{|\omega|\eexp{i\omega\tau}}{\omega+i\epsilon}\AB_{-\omega}
\eeq
The $i\epsilon$ is irrelevant since $|\omega|\delta(\omega)=0$, hence $\ab(\tau)$ is real.
The weight of each instanton is $\Lambda\Delta\sim\eexp{-S_{instanton}}$ with $S_{instanton}\sim \sqrt{g_2}$, $\Lambda$ is present so that $\Delta$ is dimensionless. The partition sum becomes
(up to a constant $\sim \int_\omega \ln (1-\alpha_\omega)$)
\beq{130}
Z_1&=&\bar Z \sum_n^\infty \sum_{\RR_\alpha}\int_{\tau_1<\tau_2<...<\tau_n}(\Lambda\Delta)^n\eexp{-S_0^{(n)}+i2\pi\sum_\alpha \RR_\alpha\cdot\ab(\tau_\alpha)}\nonumber\\
 \bar Z&=&\int {\cal D}\bar C_\tau\,\eexp{-\half N\int_\omega \frac{|\omega|}{2\pi K}|\bar C_\omega|^2+\int_\omega\frac{|\omega|}{1-\alpha_\omega} \bar C_\omega\sum_i A^i_{-\omega}}
\eeq
where $\bar Z$ is the $\bar C_\tau$ dependent part and $S_0^{(n)}$ is the Gaussian part
 $\int_\omega \frac{|\omega|}{2\pi K}\tilde\CC_\omega\cdot\tilde\CC_{-\omega}$ with n instantons inserted (a term $\sim\sum_\alpha\RR_\alpha=0$ by boundary condition is added),
\beq{131}
S_0^{(n)}=\half\int_\omega \frac{|\omega|}{2\pi K}\sum_{\alpha,\beta}4\pi^2\RR_\alpha\cdot\RR_\beta
\frac{\eexp{i\omega(\tau_\alpha-\tau_\beta)}-1}{\omega^2+\epsilon^2}=
-\frac{1}{ K}\sum_{\alpha\neq\beta}\RR_\alpha\cdot\RR_\beta\ln \Lambda|\tau_\alpha-\tau_\beta|
\eeq
The partition sum is therefore, using $\int\frac{\eexp{i\omega \tau}-1}{|\omega|}\frac{d\omega}{2\pi}=2\int_{1/|\tau|}^\Lambda\frac{-1}{\omega}\frac{d\omega}{2\pi}
=\frac{-1}{\pi}\ln \Lambda |\tau|$ with a high frequency cutoff $\Lambda$,
\beq{132}
Z_1=\bar Z \sum_{n=0}^\infty\sum_{\{\RR_\alpha\}}\int_{\tau_1,...,\tau_n}(\Lambda\Delta)^n\,\eexp{\frac{1}{ K}\sum_{\alpha\neq\beta}\RR_\alpha\cdot\RR_\beta\ln \Lambda|\tau_\alpha-\tau_\beta|+i2\pi\sum_{\alpha}\RR_\alpha\cdot \ab(\tau_\alpha)}
\eeq
Using $B^i_\omega=C^i_\omega+\frac{\alpha_\omega}{1-\alpha_\omega}\bar C_\omega=\tilde C^i_\omega+\frac{1}{1-\alpha_\omega}\bar C_\omega$ we find that for $\Delta=0$
\beq{133}
\langle C^i_\omega C^j_{-\omega}\rangle&=&\langle |\bar C_\omega|^2\rangle=\frac{2\pi K}{TN|\omega|}\nonumber\\
\langle B^i_\omega B^j_{-\omega}\rangle&=&\frac{1}{(1-\alpha_\omega)^2}\langle |\bar C_\omega|^2\rangle=\frac{N_\omega}{TNM\omega^2}
\eeq
This result can be obtained also with the original $B^i_\tau$ variables, the $C^i_\tau$ variables are needed only for the following duality.

Consider now a dual action, defined on the lattice $\RR$ with a vector field $\theta_i(\tau)$
\beq{134}
S_2=\half\int_\omega \frac{K|\omega|}{2\pi}|\bm\theta(\omega)|^2-\Lambda\Delta\sum_\RR\int_\tau \eexp{i\RR\cdot({\bm \theta}(\tau)+ 2\pi\ab(\tau))}
\eeq
The dual partition sum is expanded, the time integrals in n-th order can be ordered so that $\frac{1}{n!}\int_{\tau_1,\tau_2,...,\tau_n}...=\int_{\tau_1<\tau_2<...<\tau_n}$, hence averaging on the $|\bm\theta(\omega)|^2$ term yields
\beq{135}
Z_2=Z_0\sum_{n=0}^\infty (\Lambda\Delta)^n\int_{\tau_1<\tau_2<...<\tau_n}\sum_{\{\RR_\alpha\}}\langle\eexp{i\sum_\alpha \RR_\alpha
\cdot{\bm \theta}(\tau_\alpha)}\rangle_0\,\eexp{i2\pi \sum_\alpha\RR_\alpha\cdot\ab(\tau_\alpha)}
\eeq
where $Z_0=\int{\cal D}{\bm\theta}\,\eexp{-\half\int_\omega \frac{K|\omega|}{2\pi}|\bm\theta(\omega)|^2}$.
Each factor $\eexp{i R_\alpha^i\theta_i(\tau_\alpha)}$ must multiply a term with $R_\alpha^i\rightarrow -R_\alpha^i$, otherwise the average diverges and its exponent vanishes; hence $\sum_\alpha\RR_\alpha=0$ and a constant -1 can be added below.
The Gaussian average involves
\beq{136}
&&\half \langle (\sum_\alpha \RR_\alpha\cdot{\bm \theta}(\tau_\alpha))^2\rangle=
\half\sum_{\alpha,\beta}R_\alpha^iR_\beta^j\langle\theta^i(\tau_\alpha)\theta^j(\tau_\beta)\rangle_0
= \half\sum_{\alpha,\beta}\RR_\alpha\cdot\RR_\beta\int_\omega
\frac{\eexp{i\omega(\tau_\alpha-\tau_\beta)}-1}{|\omega|K/2\pi}\nonumber\\
&&=-\frac{1}{K}\sum_{\alpha\neq\beta}\RR_\alpha\cdot\RR_\beta\ln \Lambda|\tau_\alpha-\tau_\beta|
\eeq
so that
\beq{137}
Z_2=Z_0\sum_{n=0}^\infty (\Lambda\Delta)^n\int_{\tau_1<\tau_2<...<\tau_n}\sum_{\{\RR_\alpha\}}
\eexp{\frac{1}{K}\sum_{\alpha\neq\beta}\RR_\alpha\cdot\RR_\beta\ln \Lambda|\tau_\alpha-\tau_\beta|+i2\pi \sum_\alpha\RR_\alpha\cdot\ab(\tau_\alpha)}
\eeq
This is identical to $Z_1$, apart from the  $\sim|\bar C_\omega|^2$ term, i.e. $Z_1(A)=\bar Z(A)Z_2(A)/Z_0$, displaying now the $\AB_\omega$ dependence. This  proves the duality.

Let us now obtain the correlation functions. Since $\langle \tilde C^i_\omega \bar C_{-\omega}\rangle=0$ the original $ B_i$ correlation is now, using $\delta a^j(\tau)/\delta A^i_\omega=-T \text{sign} \omega \,\eexp{-i\omega \tau}\delta_{i,j}$
\beq{138}
&&\langle  B^i_\omega  B^j_{-\omega}\rangle=\frac{1}{Z_1}\frac{1}{T^2\omega^2}\frac{\delta^2 Z_1}{\delta A^i_{-\omega}\delta A^j_{\omega}}|_{A=0}=\frac{1}{\bar Z}\frac{1}{T^2\omega^2}\frac{\delta^2 \bar Z}{\delta A^i_{-\omega}\delta A^j_{\omega}}|_{A=0}+\frac{1}{Z_2}\frac{1}{T^2\omega^2}\frac{\delta^2 Z_2}{\delta A^i_{-\omega}\delta A^j_{\omega}}|_{A=0}\nonumber\\&&=\frac{1}{(1-\alpha_\omega)^2}\langle |\bar C_\omega|^2\rangle+\frac{i2\pi\text{sign}\omega }{Z_2 T\omega^2}\frac{\delta}{\delta A^j_{\omega}}\int{\cal D}{\bm\theta} \sum_{\RR}\Lambda\Delta\int_{\tau}\eexp{i\RR\cdot({\bm \theta}_{\tau}+2\pi \ab_{\tau})}R_i\eexp{i\omega \tau}\eexp{-S_2}|_{A=0}\\
&&=\frac{N_\omega}{TNM\omega^2}+\frac{4\pi^2}{\omega^2}\sum_{\RR,\RR'}(\Lambda\Delta)^2
\int_{\tau,\tau'}\langle \eexp{i\RR\cdot{\bm \theta}_\tau+i\RR'\cdot{\bm \theta}_{\tau'}}\rangle R_iR_j'\eexp{i\omega(\tau-\tau')}
+\Lambda\Delta\frac{(2\pi)^2}{\omega^2}\sum_\RR R^iR^j\int_\tau \langle\eexp{i\RR\cdot{\bm\theta}_\tau}\rangle\nonumber
\eeq
This is an exact relation, which we now use for perturbative expansion. To 2nd order, similar to Eq. (\ref{e120}),
\beq{139}
&&\langle\eexp{i\RR\cdot(\bm\theta_\tau-\bm\theta_{\tau'})}\rangle=\eexp{-\half \RR^2\int_\omega\frac{4\pi}{K|\omega|}(1-\cos\omega(\tau-\tau'))}=
(\Lambda|\tau-\tau'|)^{-\frac{2}{K}(1-\frac{1}{N})}\qquad  \Rightarrow\nonumber\\
&&T\langle  B^i_\omega  B^j_{-\omega}\rangle=\frac{N_\omega}{NM\omega^2}+\frac{4\pi^2}{\omega^2}(\Lambda\Delta)^2\sum_\RR R^iR^j
\int_{\tau>1/\Lambda} (\Lambda\tau)^{-\frac{2}{K}(1-\frac{1}{N})}(1-\eexp{i\omega\tau})\nonumber\\
&& =\frac{N_\omega}{NM\omega^2}+\frac{4\pi^2\Delta^2}{|\omega|}\sum_\RR R^iR^j\frac{1}{\frac{2}{K}(1-\frac{1}{N})-1}
\left(\frac{\omega}{\Lambda}\right)^{\frac{2}{K}(1-\frac{1}{N})-2}
\eeq
Interpreting $\omega=\Lambda'$ as a new cutoff we can use the RG, Eq. \eqref{e142} below so that
$\Delta\left(\frac{\omega}{\Lambda}\right)^{\frac{1}{K}(1-\frac{1}{N})-1}\rightarrow
\Delta[1-(\frac{1}{K}(1-\frac{1}{N})-1)\frac{d\Lambda}{\Lambda}=\Delta^R$, so that near the critical point $K_c=1-\frac{1}{N}$
\beq{140}
 T\langle  B^i_\omega  B^j_{-\omega}\rangle=\frac{N_\omega}{NM\omega^2}+\frac{4\pi^2}{|\omega|}
\sum_\RR R^iR^j (\Delta^R)^2=\frac{N_\omega}{NM\omega^2}+\frac{4\pi^2}{|\omega|}
2(\delta_{i,j}-\frac{1}{N}) (\Delta^R)^2
\eeq
The sum on $\RR$ is performed using $(R_\nu)^i=\frac{1}{N}-\delta_{\nu i} \,(a=1,2...,N)$ as well as the $-(R_\nu)^i$ vectors, hence $\sum_\RR R^iR^j=2(\delta_{ij}-\frac{1}{N})$.
From the critical behavior below \eqref{e142} the correction term is $\sim(K_c-K)^2$ for $N=3$ while it is $\sim K_c-K$ for $N\geq 4$.

An alternative form of the action can be obtained by shifting in \eqref{e134} ${\bm \theta}(\tau)\rightarrow {\bm \theta}(\tau)-2\pi \ab(\tau)$ ($\ab(\tau)$ should be also periodic) and using $\delta a^i_\omega/\delta A^j_\omega=-\text{sign}\omega\delta_{i,j}$,
\beq{141}
&&S_2=\half\int_\omega \frac{K|\omega|}{2\pi}|\bm\theta_\omega-2\pi \ab_\omega|^2-\Lambda\Delta\sum_\RR\int_\tau \eexp{i\RR\cdot{\bm \theta}(\tau)}\\
&&\langle  B^i_\omega B^j_{-\omega}\rangle=\frac{N_\omega}{TNM\omega^2}+
\frac{K}{TZ_2\omega}\int_{\bm\theta}\frac{\delta}{\delta A^j_{\omega}}(\theta^i_{\omega}-2\pi a^i_\omega)\eexp{-S_2}=\frac{N_\omega}{TNM\omega^2}+\frac{2\pi K}{T|\omega|} \delta_{i,j}
-K^2\langle \theta^i_\omega\theta^j_{-\omega}\rangle|_{A=0}\nonumber
\eeq
For $\Delta=0$ this indeed reproduces (\ref{e133}).

Before proceeding to self duality, we consider RG at small $\Delta$. For each $\RR$ term
\beq{142}
&&\Delta^R\Lambda'=\Delta\Lambda\langle\eexp{i\RR\cdot{\bm \theta}(\tau)}\rangle=\Delta\Lambda\eexp{-\half \sum_{i,j}R_iR_j\langle\theta_i(\tau)\theta_j(\tau)\rangle}=
\Delta\Lambda\eexp{-\RR^2\int_{\Lambda'}^\Lambda\frac{d\omega}{K\omega}}
=\Delta\Lambda(1-\frac{\RR^2}{K}\ln\frac{\Lambda}{\Lambda'})\nonumber\\
&&\Delta^R=\Delta[1+(1-\frac{1}{K}(1-\frac{1}{N})\frac{d\Lambda}{\Lambda}]
\eeq
The fixed point is at $K_c=1-\frac{1}{N}$. For $N\geq 4$ there are no $\Delta^2$ terms since $\RR\pm\RR'$ are longer minimal length vectors (unlike the $\VV$ vectors) and are therefore less relevant.
The next order is $O(\Delta^3)$ with a $1/N$ coefficient for large N (Ref. \onlinecite{yi2} table I),
hence the phase boundary is $\Delta_c\sim\sqrt{N}\sqrt{K_c-K}$
approaching steeply the $\Delta=0$ point.\\

\subsection{A. Self Duality}

Here we derive the self-dual relations given in the main text.
For $N=2,3$ the symmetry of the $\VV$ and $\RR$ lattices are the same, only the length of these vectors differ. For $N=2$, $||\VV||/||\RR||=\sqrt{2}/\sqrt{\half}=2$. Therefore, rescaling $\bm\theta\rightarrow 2\bm\theta$ and comparing with $S_1$ Eq. (\ref{e126}) at $\AB=0$, and using both (\ref{e125},\ref{e141}) yields
\beq{143}
&&S_2^{N=2}=\frac{1}{2} \int_\omega \frac{K|\omega|}{2\pi}|\bm\theta_\omega|^2-\Lambda\Delta\sum_\VV\int_\tau\eexp{i\VV\cdot\bm\theta_\tau/2}
\,\Rightarrow\qquad
\langle \theta^i_\omega \theta^j_{-\omega}\rangle=4\langle C^i_\omega C^j_{-\omega}\rangle_{K\rightarrow 1/(4K),g_2\rightarrow\Delta}=\nonumber\\ && 4[\langle B^i_\omega B^j_{-\omega}\rangle_{1/(4K),\Delta}-\frac{1}{TM\omega^2}]=\frac{1}{K^2}[\frac{N_\omega}{TNM\omega^2}
+\frac{2\pi K}{T|\omega|}\delta_{i,j}-\langle B^i_\omega B^j_{-\omega}\rangle_{K,g_2}]
\eeq
The self dual point is at $K=\half,\,\Delta=g_2$ and the last relation determines the correlation at this point as the average of the $g_2=0$ (Eq. \ref{e117}) and $\Delta=0$ (Eq. \ref{e133}) cases
i.e. $T \langle B^i_\omega B^j_{-\omega}\rangle_{1/2,g_2=\Delta}=\half(D_{i,j}+\frac{N_\omega}{NM\omega^2})$ . Since the transition line is vertical and the $\Delta(g_2)$ relation is not known precisely this is not such a useful information.

For N=3 the $\RR$ lattice is also triangular with $||\VV||/||\RR||=\sqrt{2}/\sqrt{2/3}=\sqrt{3}$.  Rescaling ${\bm \theta}(\tau)\rightarrow\sqrt{3}\bm\theta$ and using both (\ref{e125},\ref{e141}) yields
\beq{144}
&&S_2^{N=3}=\half\int_\omega \frac{K|\omega|}{2\pi}|\bm\theta_\omega|^2-\Delta\sum_\VV\int_\tau \eexp{i\VV\cdot{\bm \theta_\tau/\sqrt{3}}} \,\Rightarrow\qquad
\langle \theta^i_\omega \theta^j_{-\omega}\rangle=3\langle C^i_\omega C^j_{-\omega}\rangle_{K\rightarrow 1/(3K),g_2\rightarrow\Delta}=\nonumber\\&&3[\langle B^i_\omega B^j_{-\omega}\rangle_{1/(3K),\Delta}-\frac{1}{TM\omega^2}]=\frac{1}{K^2}[\frac{N_\omega}{TNM\omega^2}
+\frac{2\pi K}{T|\omega|}\delta_{i,j}-\langle B^i_\omega B^j_{-\omega}\rangle_{K,g_2}]
\eeq
 The self-dual point is at $K=\frac{1}{\sqrt{3}},\,\Delta=g_2$ where the correlation becomes
\beq{145}
K=\frac{1}{\sqrt{3}}:\, \langle  B^i_\omega B^j_{-\omega}\rangle=\half\{\frac{N_\omega}{TNM\omega^2}
+\frac{2\pi K}{T|\omega|}\delta_{i,j}+\frac{1}{TM\omega^2}\}
\eeq
which is Eq. (24) of the main text, and again is precisely the average of the $g_2=0$ (Eq. \ref{e118}) and $\Delta=0$ (Eq. \ref{e133}) cases.
The result is also consistent with the sum rule $T\sum_i\langle  B^i_\omega B^j_{-\omega}\rangle=\sum_iD_{i,j}=\frac{N_\omega}{M\omega^2}$.

 \section{IV. Conductance}
 In this section we derive the conductance of LL wires with normal leads, given in the text in Eq. (22).
 Our surmise is that the resistance in the strongly coupled phase is the sum of individual resistances. The reasoning is that an equal current $I$ flows in all wires, hence the resistance $R_i$ of each wire (actually a contact resistance) leads to a total dissipation equal to $\sum_i I^2R_i$. Since the voltage is applied on one wire its effective resistance is $R_{eff}=\sum_iR_i$. The equality of all currents implies also that all components of the conductance matrix $G_{ij}$ are equal. In the following we confirm this surmise in various geometries. We start, however, with deriving a local conductance, conceptually important for the following derivation,
and given in the text in Eq. (24).

\subsection{A. Local Conductance}

Consider the local conductance \cite{giamarchi}, i.e. the current at $x=0$ on chain $i$ in response to a voltage in the range $(-L/2,\,L/2)$ on chain j
\beq{146}
G^{\rm local}_{ij}(\omega)=-\frac{e^2}{\pi^2 \hbar}i(\omega+i\delta)\frac{1}{L}\int_{-L/2}^{L/2}dx'\langle
\phi_i(x=0,\omega_n)\phi_j(x',-\omega_n)\rangle|_{i\omega_n\rightarrow \omega+i\delta}
\eeq
As shown in the next subsection the response in the DC limit $\omega \to 0$ is $x,\,x'$ independent, hence we can take $L\rightarrow 0$ so that the conductance is expressed in terms of the local
fields at $x=0$, $\phi_i(\omega_n)$, of our nonlinear problem. In terms of the fields $B^i_\omega$
\beq{147}
\phi^i_\omega&=&\half(B^i_\omega+X_\omega)=\half(B^i_\omega-\frac{1}{N_\omega}\sum_jB^j_\omega+\tilde X_\omega)
\eeq
and using the sum rule \eqref{e116} we obtain
\beq{148}
G^{\rm local}_{ij}(\omega)=\frac{e^2}{2\pi h}\omega[T\langle B^i_{\omega_n} B^j_{-\omega_n}\rangle-\frac{1}{M\omega_n^2}]
|_{i\omega_n\rightarrow \omega+i\delta}
\eeq
Using the analytic continuation
\beq{149}
\text{sign}(\omega_n)=\int \frac{d\epsilon}{\pi}\frac{i}{\epsilon +i\omega_n}\rightarrow
\int \frac{d\epsilon}{\pi}\frac{i}{\epsilon +\omega +i\delta}=+1
\eeq
and Eqs. (\ref{e121},\ref{e140}),
we obtain the conductance in the vicinity of the fixed lines ($g_2=0$ and
$\Delta=0$) associated to the two phases, as well as at the self dual point
\beq{150}
g_2 \,\, \text{small} &:&\qquad  G^{\rm local}_{i,j}(\omega)=\frac{e^2}{h}[K\delta_{i,j}+2\pi(g_2^R)^2(1-N\delta_{i,j})]\nonumber\\
\Delta \,\, \text{small} &:& \qquad  G^{\rm local}_{i,j}(\omega)=\frac{e^2}{Nh}[K-4\pi (\Delta^R)^2 (1-N\delta_{i,j})]\nonumber\\
\mbox{self-dual}&:&\qquad  G^{\rm local}_{i,j}(\omega)=\half\frac{e^2}{h}K[\delta_{i,j}+\frac{1}{N}]
\qquad (N=2,3)
\eeq
 Inserting the RG fixed point values in these equations
also give the leading corrections of the conductance along the phase transition line, as discussed
in the text. The conductance matrix thus varies continuously along the phase transition line.

\subsection{B. Local Conductance for the Superconducting Grain System}

In this subsection the fields refer to the charge sector \cite{giamarchi}, i.e. $\phi,\theta\rightarrow \phi_\rho,\theta_\rho$; also the TLL index $i=1,2,...,N$ is implicit.
The superconducting grain realization of our model corresponds to the Hamiltonian
\beq{201}
{\cal H}_2=\frac{1}{2\pi}\sum_i\int_x[K_\rho(\nabla\theta)^2+
\frac{1}{K_\rho}(\nabla\phi)^2]-g\sum_i\cos[X-\sqrt{2}\theta(0)]+E_cP^2
\eeq
where $P$ is conjugate to the superconductor's phase $X$ and $E_c$ is the charging energy. The $\sqrt{2}$ factor corresponds to forming a singlet pair on each TLL that can Josephson tunnel into the s-wave superconductor; we assume repulsive backscattering on the TLLs so that the pairing coefficient has $\langle\cos\sqrt{2}\phi_\sigma\rangle\neq 0$. The current operator includes now a term $\sim\sin[X-\sqrt{2}\theta(0)]\text{sign}(x)$ that makes the previous formalism for evaluating linear response to a vector potential inconvenient. We proceed instead by developing a Kubo formula as a linear response to a scalar potential. Since we are interested in the local response to a field localized at $x=0$ we can use a potential $V(x,t)=V\eexp{-i\omega t}\theta(-x)$ so that the perturbation to the Hamiltonian is
\beq{202}
{\cal H}^{ex}=\int_x\rho(x)V(x,t)=\frac{\sqrt{2}}{\pi}\int_x\nabla\phi(x)V(x,t)
=\frac{\sqrt{2}}{\pi}\phi(0)V\eexp{-i\omega t}
\eeq
The current $j(x,t)=\frac{\sqrt{2}}{\pi}\dot\phi(x,t)$ as a linear response to this perturbation identifies the conductance, which in Matsubara frequency $\omega_n$ is
\beq{203}
G(x,\omega_n)=-\frac{2\omega_n}{\pi^2}\langle\phi_n(x)\phi_{-n}(0)\rangle
\eeq
This form was also given by Ref. \onlinecite{mora}. The action in terms of both $\phi_n(x),\theta_n(x)$ can be written as
\beq{204}
S=&&\frac{1}{2\pi}\sum_{n}\int_x\{ K_\rho|\nabla\theta_n|^2+\frac{1}{K_\rho}
(\nabla\phi_n-K_\rho\omega_n\theta_n)(\nabla\phi_{-n}+K_\rho\omega_n\theta_{-n})
+ K_\rho\omega_n^2|\theta_n|^2\}\nonumber\\&& -g\int_\tau\{\cos[X_\tau-\sqrt{2}\theta(0,\tau)]+\frac{\dot X^2}{4E_c}\}
\eeq
Since the action is Gaussian in $\theta_n^i(x\neq 0)$ it is useful to separate the saddle point solution for $\theta_n(x)$, as well as shift the field $\phi_n(x)$,
\beq{205}
\theta_n(x)=\theta_n(0)\eexp{-|\omega_n x|}+\delta\theta_n(x),\qquad \nabla\tilde\phi_n=\nabla\phi_n-K_\rho\omega_n\theta_n
\eeq
so that the action becomes
\beq{206}
S=&&\frac{1}{2\pi}\sum_{n}\{ \int_x[ K_\rho\omega_n^2|\delta\theta_n(x)|^2+\frac{1}{K_\rho}
|\nabla\tilde\phi_n(x)|^2]+K_\rho|\omega_n||\theta_n(0)|^2\}\nonumber\\&&
-g\int_\tau\{\cos[X_\tau-\sqrt{2}\theta(0,\tau)]+\frac{\dot X^2}{4E_c}\}
\eeq
Rewriting the conductance \eqref{e203} with the shifted variables \eqref{e205} the Gaussian fluctuations of $\tilde\phi_n(0)$ and of $\delta\theta_n(x)$ cancel, hence
\beq{207}
G^{ij}(\omega_n)=\frac{2e^2}{\pi^2}\omega_n K_\rho^2\langle\theta_n^i(0)\theta_{-n}^j(0)\rangle
\eeq
restoring here the TLL indices $i,j$.

The $\theta_n(0)$ dependent action \eqref{e206} is identical to the $\phi_n(0)$ dependent action in Eq. (6) of the main text with the replacement $\theta_n^i\rightarrow\sqrt{2}\theta_n^i$, hence $K\rightarrow \frac{1}{2K_\rho}$. The conductance \eqref{e207} of the superconducting grain is thus obtained from Eq. (24) of the main text with the replacement $K\rightarrow 4K_\rho^2/(2K_\rho)=2K_\rho$. In weak coupling $g=0$ this is the expected result, with the factor 2 corresponding this spinfull case. In the strong coupling limit one needs the mapping \eqref{e207} to realize that the currents on all TLLs are equal and then the argument above Eq. (22) in the main text yields the expected conductance.

\subsection{C. Normal leads for N TLLs}

We consider here the strongly coupled phase of N TLLs with boundary conditions of normal leads on each TLL, represented by LLs with Luttinger parameter $K_L$ (eventually $K_L=1$). In this geometry one can measure current and voltage at either of the external leads. For $N=1$ the conductance is \cite{safi,maslov,maslov2,ponomarenko2} $\frac{e^2}{h}K_L\rightarrow \frac{e^2}{h}$.

In view of the space dependence our duality method is not feasible, hence we replace the nonlinear coupling in the action Eq. (13) of the main text by a Gaussian one $\half\mu\sum_{ij}[\phi_i(0,\tau)-\phi_j(0,\tau)]^2$ . This replacement
is justified because (i) all $G_{ij}$ become equal, indicating equal currents as needed, (ii) the result is independent of the coupling $\mu$ in the limit $\omega_n/\mu\rightarrow 0$, (iii) the result reproduces the local conductance Eq. (\ref{e150}) in the formal limit $K_L\rightarrow K$. In the latter limit the TLL is extended without leads, corresponding to local conductance, which in turn incorporates the full nonlinearity of our system.
  We note that a similar replacement was also done in Ref. \onlinecite{chou}. The LL action, allowing for space dependence of $u(x),K(x)$, becomes
\beq{151}
S=&&\half \int_{\omega_n}\int_x\left\{\sum_i[\frac{\omega_n^2}{u(x)K(x)}|\phi_i(\omega_n,x)|^2+\frac{u(x)}{K(x)}
|\partial_x\phi_i(x)|^2]
+\mu\sum_{i\neq j}\delta(x)|\phi_i(\omega_n,x))-\phi_j(\omega_n,x)|^2\right\}\nonumber\\
&&\equiv \int_{\omega_n}\int_x \phi (\omega_n, x)
C^{-1}_{i,j}(\omega_n;x,x')\phi^*(\omega_n,x')
\eeq
This identifies the retarded correlation $C^{-1}_{ik}(\omega_n,x,x')$.  We can then define the
conductance as
\newline
$G_{i,j}(\omega_n,x,x')=\frac{e^2\omega_n}{\pi^2\hbar}C_{i,j}(\omega_n,x,x')|_{i\omega_n\rightarrow \omega+i\delta}$,
where the $x,x'$ dependence actually drops out in the DC limit (see below).
Upon multiplication of $C^{-1}_{i,k}(\omega_n;x,x'')$ on the right by $C_{kj}(\omega_n,x'',x')$ and summation on $k,\,x''$
\beq{152}
\sum_k&&\left\{[\frac{\omega_n^2}{u(x)K(x)}-\partial_x\frac{u(x)}{K(x)}
\partial_x+\mu(N-1)\delta(x)]\delta_{i,k}-\mu\delta(x)(1-\delta_{i,k})\right\}C_{kj}(\omega_n,x,x')
=\pi\delta(x-x')\delta_{i,j}
\eeq
The symmetry among the wires and the common boundary conditions allow for all the diagonal $C_{ii}$ to be equal and also all off diagonal $C_{i\neq j}$ to be equal. Hence two equations of motion
\beq{153}
&&[\frac{\omega_n^2}{u(x)K(x)}-\partial_x\frac{u(x)}{K(x)}
\partial_x+M(N-1)\delta(x)]C_{ii}-M(N-1)\delta(x)C_{i\neq j}=\pi\delta(x-x')\nonumber\\
&& [\frac{\omega_n^2}{u(x)K(x)}-\partial_x\frac{u(x)}{K(x)}
\partial_x+M\delta(x)]C_{i\neq j} -M\delta(x)C_{ii}=0
\eeq
The correlation functions need outgoing waves at the boundary as appropriate for normal wires. This also implies dissipation, e.g. at $x>\half L$ and $\nu\equiv |\omega_n|\rightarrow -i(\omega+i\delta)$, then $\eexp{-\nu x/u_L}\rightarrow \eexp{i(\omega+i\delta)x/u_L}$. The solution can be written in the following form
\beq{154}
C_{ii}= && A\eexp{\nu x/u_L} \qquad \qquad \qquad\qquad x<-\half L \nonumber\\
  && B\eexp{\nu x/u}+C\eexp{-\nu x/u}  \qquad x<x'<0; \, x<0<x' \nonumber\\
  && B'\eexp{\nu x/u}+C'\eexp{-\nu x/u}  \qquad 0<x<x' \nonumber\\
  && D\eexp{\nu x/u}+E\eexp{-\nu x/u}  \qquad  0>x>x' \nonumber\\
  && D'\eexp{\nu x/u}+E'\eexp{-\nu x/u}  \qquad x>x'>0; \, x>0>x' \nonumber\\
  && F\eexp{-\nu x/u_L} \qquad \qquad \qquad\qquad x>\half L\nonumber\\
C_{i\neq j}=&& a\eexp{\nu x/u_L}\qquad\qquad\qquad\qquad  x<-\half L\nonumber\\
&& b\eexp{\nu x/u}+c\eexp{-\nu x/u}\qquad\qquad   -\half L<x<0 \nonumber\\
&& d\eexp{\nu x/u}+e\eexp{-\nu x/u} \qquad\qquad  0<x<\half L\nonumber\\
&& f\eexp{-\nu x/u_L}\qquad\qquad\qquad\qquad x>\half L
\eeq
 where the coefficients are implicit functions of $\nu,\,x',L$,
which are determined by the conditions of continuity and derivative
jumps at $x=0,\,x=x', x= \pm L/2$. This leads to a complicated expression
which simplifies drastically in the limit $\nu\rightarrow 0$ at fixed $x,x',L$
leading to
  \beq{155}
  B+C \simeq A \simeq \frac{\pi K_L}{2\nu N}+O(1)\nonumber\\
  \Rightarrow  G_{ij}=\frac{e^2}{h}\frac{K_L}{N}\rightarrow\frac{e^2}{h}\frac{1}{N}
 \eeq
 This result coincides with that in Eq. (28) in the main text on the $\Delta=0$ fixed line if $K_L\rightarrow K$ instead of
 $K_L\rightarrow 1$. This Eq. (24) was derived with the full nonlinear coupling using duality, hence this coincidence supports the replacement of the nonlinear coupling by a Gaussian term for the purpose of evaluating the DC conductance. The conductance at the fully coupled fixed point $\Delta=0$,
 with normal leads, is then $G_{i.j}=\frac{e^2}{Nh}$ and is consistent with our surmise.\\

It is also possible to calculate the {\it non-local conductance}, i.e. the response of the current at position $x$ to a voltage at position $x'\neq x$, using the above equations.
A similar quantity was defined and obtained in Ref. \onlinecite{safi} for a single LL, although calculated with a different method. Following Ref. \onlinecite{safi} we take $x=-\half L$ and $x'=\half L$, so that resonances are expected at the eigenfrequencies of the wire, $\frac{2\pi L}{u}\times$integer.
This response is defined as
\beq{156}
G_{ij}^{\rm non-local}(\omega,L) =
\frac{e^2 \omega_n}{\pi^2 \hbar} C_{ij}(\omega_n;x=- \frac{L}{2},x'=\frac{L}{2}) |_{i\omega_n\rightarrow \omega+i\delta}
\eeq
Keeping $L$ and $\nu$ finite in the above calculation, and then taking $L$ large with
the product $\nu L$ finite, we obtain
\beq{157}
G_{ij}^{\rm non-local}(\omega,L) = \frac{e^2}{h N} F(\omega L/u)
\eeq
where $F(z)$ and its real part are periodic functions of $z$
\beq{158}
F(z) = \frac{2 i K K_L^2}{\left(K^2+K_L^2\right) \sin (z)+2 i K K_L \cos (z)}
\quad , \quad {\rm Re} F(z) = \frac{1}{\frac{\left(K^2+K_L^2\right){}^2 \sin (z) \tan (z)}{4 K^2 K_L^3}+\frac{\cos
   (z)}{K_L}}
\eeq
with $F(z)=K_L-\frac{z^2 \left(K^4+K_L^4\right)}{4 \left(K^2 K_L\right)}+O\left(z^3\right)$
at small $z$. The function $F(z)$ coincides with the result of Ref. \onlinecite{safi} for $K_L=1$.
The dependence on $N$ of the non-local conductance in the coupled phase is thus very simple. While the DC limit is independent of $K$ a measurement of the whole response $F(z)$ can determine the interaction parameter $K$.

 \subsection{D. One wire and one loop}

Consider now $N=2$ with the same coupling between wires 1 and 2 as in Eq. \eqref{e151}, and assuming for generality that the 2nd wire has LL parameters $u',K'$. Leads are attached only to wire 1 at $x=\pm\half L$ while wire 2 has uniform $u',K'$ and uniform boundary conditions at
 $x=\pm\half L'$. The voltage is applied at $|x|<\half L$ and we assume that $L\leq L'$, a slightly simpler case. We extend the range that $x'$ is needed to $|x'|<\half L'$ so that the contact points of wire 1 are included.
The correlations $C_{ij}(\omega_n,x,x')$ satisfy
\beq{159}
\left(\begin{array}{cc}\frac{\omega_n^2}{u(x)K(x)}-\partial_x(\frac{u(x)}{K(x)}\partial_x)+M\delta(x) & -M\delta(x)\\-M\delta(x) & \frac{\omega_n^2}{u'K'}-\frac{u'}{K'}\partial^2_x+M\delta(x)\end{array}\right)
\left(\begin{array}{cc} C_{11} & C_{12} \\ C_{21} & C_{22}\end{array}\right)=\pi\delta(x-x')\,{\bf 1}
\nonumber\\
\eeq
The correlation functions $C_{11}(\omega_n,x,x')$ and $C_{12}(\omega_n,x,x')$ obey, in the variable $x$, the boundary conditions of wire 1 while $C_{21}(\omega_n,x,x')$ and $C_{22}(\omega_n,x,x')$ obey the periodic boundary conditions of wire 2.
Therefore, the solutions have the form
\beq{160}
C_{11}= && A_1\eexp{\nu x/u_L} \qquad \qquad \qquad\qquad x<-\half L \nonumber\\
  && B_1\eexp{\nu x/u}+C_1\eexp{-\nu x/u}  \qquad \qquad x<x'<0; \, x<0<x' \nonumber\\
  && B'_1\eexp{\nu x/u}+C'_1\eexp{-\nu x/u}  \qquad\qquad 0<x<x' \nonumber\\
  && D_1\eexp{\nu x/u}+E_1\eexp{-\nu x/u}  \qquad\qquad  0>x>x' \nonumber\\
  && D'_1\eexp{\nu x/u}+E'_1\eexp{-\nu x/u}  \qquad\qquad x>x'>0; \, x>0>x' \nonumber\\
  && F_1\eexp{-\nu x/u_L} \qquad \qquad \qquad \qquad x>\half L\nonumber\\
 C_{12}=&& a_1\eexp{\nu x/u_L}\qquad\qquad\qquad\qquad  x<-\half L\nonumber\\
&& b_1\eexp{\nu x/u}+c_1\eexp{-\nu x/u}\qquad\qquad  -\half L<x<0 \nonumber\\
&& d_1\eexp{\nu x/u}+e_1\eexp{-\nu x/u} \qquad\qquad  0<x<\half L\nonumber\\
&& f_1\eexp{-\nu x/u_L}\qquad\qquad\qquad\qquad  x>\half L\nonumber\\
C_{22}=&& B_2\eexp{\nu x/u'}+C_2\eexp{-\nu x/u'} \qquad\qquad -\half L'<x<x'<0;\,-\half L'<x<0<x'\nonumber\\
&& B_2'\eexp{\nu x/u'}+C_2'\eexp{-\nu x/u'}\qquad\qquad  0<x<x'<\half L' \nonumber\\
&& D_2\eexp{\nu c/u'}+E_2\eexp{-\nu x/u}\qquad\qquad  -\half L'<x'<x<0  \nonumber\\
&& D_2'\eexp{\nu x/u'}+E_2'\eexp{-\nu x/u'}\qquad\qquad  \half L'>x>x'>0;\, \half L'>x>0>x'\nonumber\\
C_{21}=&& b_2\eexp{\nu x/u'}+c_2\eexp{-\nu x/u'}\qquad\qquad -\half L'<x<0\nonumber\\
&& b_2'\eexp{\nu x/u'}+c_2'\eexp{-\nu x/u'} \qquad\qquad  0<x<\half L'
  \eeq
The various boundary conditions of continuity and jump of derivatives lead, after some algebra, to a conductance matrix  in the DC limit $\nu \to 0$ equal to $G_{ij}=
\frac{e^2}{h}+O(\omega)$. This result is consistent with our surmise that the resistance of the coupled system equal to the sum of individual resistances. The periodic loop is ideal, hence its resistance vanishes, while the single wire with leads has the well known resistance \cite{safi,maslov1,maslov2,ponomarenko} of $\frac{e^2}{h}$.

We note that in Ref. \onlinecite{chou} this geometry is considered in the main text, while the Gaussian coupling is slightly different, i.e. it is along the whole length $L'$. In contrast to the result above they claim that the DC conductance is $K$ dependent. However, in their supplement \cite{chou} they seem to use outgoing boundary conditions for the ideal loop, i.e. not periodic boundary conditions. Outgoing boundaries of an ideal LL wire imply a conductance of $\frac{e^2}{h}K'$ (though it is not clear where dissipation originates; in any case it is not experimentally feasible) while the wire with boundaries has the usual $\frac{e^2}{h}$. Summing up resistances, according to our surmise, gives
$G_{ij}=\frac{e^2}{h}\frac{1}{1+1/K'}$, which in fact we have also evaluated directly for these boundary conditions. The latter result is also given in Ref. \onlinecite{chou}, however, it does not correspond to the realistic geometry and boundary conditions.

\section{V. 1/N expansion}

We study here the case of large $N$ using a $1/N$ expansion.
Starting from our model in the main text, Eqs. (1,3), we first derive an effective action to order $g^4N$ (Eq. \ref{e166}), extending the leading order
$\sim g^2N$ case studied in Ref. \onlinecite{bh}. We present here the mean field solution, valid near $K=\half$, and show that a phase transition survives for $N\geq 4$ (Eq. \ref{e174}).
The approach is quite different from the one in the main text, since here we first integrate over all the degrees of freedom of the LL's, seen as a bath, and
study the effective action for $X(\tau)$. While this expansion is valid for $N\gg 1$ it does capture some of the features studied in the main text, i.e. the presence of a phase transition for a finite $N$.

\subsection{A. effective action}
Consider first the linear coupling to the density on chain $i$, i.e. $-g_0\partial_x\phi_i(X_\tau,\tau)/\pi$ ($i=1,...,N$)  which can be integrated exactly to yield the long wavelength impurity action $S_{long}$,
\beq{161}
&&\langle \eexp{-g_0\int_0^{\beta}\partial_x\phi_i(X(\tau),\tau)/\pi}\rangle=\eexp{-S_{long}}\nonumber\\
&&S_{long}=-\half \frac{g^2}{\pi^2}\int_{\tau,\tau'}\langle \partial_x\phi_i(X(\tau),\tau)\partial_x\phi_i(X(\tau'),\tau')\rangle_0=
-\half g_0^2\frac{K}{2\pi}\int_{\tau,\tau'}
\frac{y_{\alpha}^2-(X(\tau)-X(\tau'))^2}{[y_{\alpha}^2+(X(\tau)-X(\tau'))^2]^2}\nonumber\\
\eeq
 where $\langle...\rangle_0$ is an average on the LL action and \cite{giamarchi}  $y_{\alpha}=\tau-\tau'+\alpha\text {sign}(\tau-\tau')$, $K$ is the luttinger parameter and $\alpha\sim 1/\rho_0$ is a cutoff. At long times the interaction is $\sim\frac{(X(\tau)-X(\tau'))^2}{y_{\alpha}^4}$, whose Fourier transform is $\sim\omega^3$, negligible relative to the mass term $M\omega^2$. This conclusion is valid for any $N\geq 1$.

Our main concern is therefore the coupling to the oscillatory term of the density Eq. (2) in the main text by expanding in $g_0$ and averaging on $\phi_i(X_\tau,\tau)$ with known LL correlations \cite{giamarchi}, i.e.
\beq{162}
\langle \eexp{2i\phi_i(X_\tau,\tau)-2i\phi_j(X_{\tau'},\tau'))}\rangle_0=
\left(\frac{\alpha^2}{y_{\alpha}^2+(X(\tau)-X(\tau'))^2}\right)^{K}\delta_{i,j}
\eeq
Already at this stage we note that we expect at long times $|X_\tau-X_{\tau'}|\ll |\tau-\tau'|$, as confirmed by the derived phase diagram \cite{bh}. Hence we neglect the $X_\tau$ terms in the correlation \eqref{e162}. The expansion in $g=\alpha_1\rho_0g_0$, equivalent to a $1/N$ expansion, then yields
\beq{163}
&&\eexp{-\half M\int_\tau \dot X(\tau)^2}\langle\eexp{-g\int_\tau\sum_i[\eexp{2i(\pi\rho_0X(\tau)-\phi_i(X(\tau),\tau))}+h.c.]}
\rangle_0=\eexp{-S_{eff}}\nonumber\\
&&S_{eff}=\half M\int_\tau \dot X(\tau)^2-\half g^2\sum_i  \int_{\tau_1,\tau_2}\langle [\eexp{i(X(\tau_1)-2\phi_i(\tau_1))}+h.c.]
[\eexp{i(X(\tau_2)-2\phi_i(\tau_2))}+h.c.]\rangle_0 -\frac{1}{4!}g^4\sum_{i_1,...,i_4}
 \int_{\tau_1,...,\tau_4} \\
 &&\langle[\eexp{i(X(\tau_1)-2\phi_{i_1}(\tau_1))}+h.c.]
[\eexp{i(X(\tau_2)-2\phi_{i_2}(\tau_2))}+h.c.][\eexp{i(X(\tau_3)-2\phi_{i_3}(\tau_3))}+h.c.]
[\eexp{i(X(\tau_4)-2\phi_{i_4}(\tau_4))}+h.c.]\rangle_{0,c}+O(g^6N)\nonumber
\eeq
where $\langle ...\rangle_{0,c}$ is a cumulant average, subtracting the next order of the $g^2$ term. In the last term we choose 2 out of 4 phases(4!/4 choices) to be positive and define them as 1,3,  (including c.c. terms), defining $\tau_{ij}=\tau_i-\tau_j$,
\beq{164}
&&S_{eff}=\half M\int_\tau \dot X(\tau)^2
-g^2N\Lambda^{-2K}\int_{\tau_1,\tau_2}\frac{\cos[X(\tau_1)-X(\tau_2)]}{|\tau_{12}|^{2K}}
+\half g^4N^2\Lambda^{-4K}\int_{\tau_1,...,\tau_4}\frac{\cos[X(\tau_1)-X(\tau_2)+X(\tau_3)-X(\tau_4)]}
{|\tau_{12}\tau_{34}|^{2K}}\nonumber\\
&&- \frac{1}{4} g^4\sum_{i_1,...,i_4}\int_{\tau_1,...,\tau_4}
\cos[X(\tau_1)-X(\tau_2)+X(\tau_3)-X(\tau_4)]\langle \eexp{2i[\phi_{i_1}(\tau_1)
-\phi_{i_2}(\tau_2)+\phi_{i_3}(\tau_3)-\phi_{i_4}(\tau_4)]}\rangle
\eeq
The second term $O(g^2N)$ is a classical long range XY model which  was studied in Ref. \onlinecite{bh}.
 We symmetrize the 3rd term, for later convenience, i.e. replace
 $|\tau_{12}\tau_{34}|^{-2K}\rightarrow\half
 [|\tau_{12}\tau_{34}|^{-2K}+|\tau_{14}\tau_{23}|^{-2K}]$.
Next we choose $i_1=i_2=1,2,...,N$ and $i_3=i_4=1,2,..,N$ but $i_1\neq i_3$ giving $N(N-1)$ terms like the last one, similarly with $i_1=i_4,\,i_2=i_3$, hence $-\half g^4[N(N-1)-N^2]=\half g^4N$.
Remaining is the term $i_1=i_2=i_3=i_4$ giving $g^4N$. Using Ref. \onlinecite{giamarchi} Eq. C.37
\beq{165}
\langle \eexp{2i[\phi(\tau_1)
-\phi(\tau_2)+\phi(\tau_3)-\phi(\tau_4)]}\rangle=\eexp{-2K[\ln\tau_{12}-\ln\tau_{13}
+\ln\tau_{14}+\ln\tau_{23}-\ln\tau_{24}+\ln\tau_{34}]}
\eeq
our final form is (with $\tau_i\rightarrow\tau_i\Lambda$)
\beq{166}
&&S_{eff}=\half M\int_\tau \dot X(\tau)^2-g^2N\Lambda^{-2K}\int_{\tau_1,\tau_2}\cos[X(\tau_1)-X(\tau_2)]\frac{1}{|\tau_{12}|^{2K}}
-\frac{1}{4} g^4N\Lambda^{-4K}\times\\
&&\int_{\tau_1,...,\tau_4} \cos[X(\tau_1)-X(\tau_2)+X(\tau_3)-X(\tau_4)]\left\{\left|\frac{\tau_{13}\tau_{24}}{\tau_{12}\tau_{14}\tau_{23}\tau_{34}}\right|^{2K}
-\frac{1}{|\tau_{12}\tau_{34}|^{2K}}--\frac{1}{|\tau_{14}\tau_{23}|^{2K}}\right\}\nonumber
\eeq

Note that the $\cos[X(\tau_1)-X(\tau_2)+X(\tau_3)-X(\tau_4)]$ term cannot be decomposed, in general, to a $\cos$ product since the $\tau$ integrals are mixed.  Note also that if the pair $\tau_1,\tau_2$ is separated by a large time $\bar T$ from the pair $\tau_3,\tau_4$ with $\bar T\gg|\tau_1-\tau_2|,|\tau_3-\tau_4|$ then
$\tau_{14},\tau_{23},\tau_{24},\tau_{13}\rightarrow \bar T$ and the first 2 terms in the last line of (\ref{e166}) cancel while the last term is small $\sim 1/{\bar T}^2$; similarly when the pair $\tau_1,\tau_4$ is well separated from the pair $\tau_2,\tau_3$, then the 1st and 3rd term cancel and the result is again $\sim 1/{\bar T}^2$. Hence well separated pairs are already included in the 1st term of $S_2$ and cancel in the 2nd order cumulant.

\subsection{B. Expansion near $K=\half$}

Near $K=\half$ the interactions in the effective action have a very long range and mean field theory is justified. We reproduce first the $Ng^2$ result \cite{bh}.
Replace each factor $\langle\eexp{iX(\tau)}\rangle=h$ as a complex order parameter, i.e. $\eexp{iX(\tau_1)-iX(\tau_2)}\rightarrow h\eexp{-iX(\tau_2)}+h^*\eexp{iX(\tau_1)}$. We now introduce the dimensionless
coupling constant, $\eta=2\pi g^2N\Lambda^{-2}$, in terms of which
\beq{167}
S_{eff}^{(1)}=\half M\int_\tau \dot X^2 -\frac{\eta\Lambda^{2-2K}}{2\pi}\left( h \int_{\tau_1,\tau_2}\eexp{-iX(\tau_2)}
\frac{1}{|\tau_{12}|^{2K}}+ h^* \int_{\tau_1,\tau_2}\eexp{iX(\tau_1)}
\frac{1}{|\tau_{12}|^{2K}}\right)
\eeq
 Expand $\eexp{-S^{(1)}_{eff}}$ to have linear term in h, the critical $\eta$ is the solution of
\beq{168}
1=\langle \eexp{i X(\tau_0)}\rangle_h/h=\frac{\eta\Lambda^{2-2K}}{2\pi}\int_{\tau_1,\tau_2} \frac{\langle \eexp{iX(\tau_0)-iX(\tau_2)}\rangle_0}{|\tau_{12}|^{2K}}=
\frac{\eta\Lambda^{2-2K}}{2\pi}\int_{\tau_1,\tau_2} \frac{\eexp{-|\tau_{02}|/2M} }{|\tau_{12}|^{2K}}=\frac{8M\eta\Lambda}{2\pi(2K-1)}
\eeq
taking the short time cutoff of $\tau_{12}$ as $1/\Lambda$.

The next order with the dimensionless $\eta_2=\frac{1}{4}g^4N\Lambda^{-4}$ becomes
\beq{169}
&&S_{eff}=S_{eff}^{(1)}+\half\eta_2\Lambda^{4-4K}\int_{\tau_1,...,\tau_4} [h\eexp{-iX(\tau_2)+iX(\tau_3)-iX(\tau_4)}+h^*\eexp{iX(\tau_1)+iX(\tau_3)-iX(\tau_4)}
\\&&+h\eexp{iX(\tau_1)-iX(\tau_2)-iX(\tau_4)}+h^*\eexp{iX(\tau_1)-X(\tau_2)+X(\tau_3)}+h.c.]
\left\{\left|\frac{\tau_{13}\tau_{24}}{\tau_{12}\tau_{14}\tau_{23}\tau_{34}}\right|^{2K}
-\frac{1}{|\tau_{12}\tau_{34}|^{2K}}-\frac{1}{|\tau_{14}\tau_{23}|^{2K}}\right\}\nonumber
\eeq
Expand $\eexp{-S_{eff}}$ to have linear term in h, only half of the 8 terms contribute,
\beq{170}
&&1=\langle \eexp{i X_0}\rangle_h/h|_{h\rightarrow 0}=\frac{8M\eta\Lambda}{2\pi(2K-1)}+\half\eta_2\Lambda^{4-4K}
\int_{\tau_1,...,\tau_4} [\eexp{iX(\tau_0)-iX(\tau_2)+iX(\tau_3)-iX(\tau_4)}\nonumber\\&&+\eexp{iX(\tau_0)-iX(\tau_1)-iX(\tau_3)+iX(\tau_4)}
+\eexp{iX(\tau_0)+iX(\tau_1)-iX(\tau_2)-iX(\tau_4)}+
\eexp{iX(\tau_0)-iX(\tau_1)+X(\tau_2)-X(\tau_3)}]\nonumber\\
&&\left\{\left|\frac{\tau_{13}\tau_{24}}{\tau_{12}\tau_{14}\tau_{23}\tau_{34}}\right|^{2K}
-\frac{1}{|\tau_{12}\tau_{34}|^{2K}}-\frac{1}{|\tau_{14}\tau_{23}|^{2K}}\right\}
\eeq
All 4 terms are identical by interchanging variables: in 2nd term 1$\leftrightarrow$2, 3$\leftrightarrow$4, in 3rd term 1$\leftrightarrow$3, in 4th term  1$\rightarrow$2, 3$\rightarrow$4, 2$\rightarrow$3, 4$\rightarrow$1.

 Performing the Gaussian averages,
\beq{171}
&&1=\frac{8M\eta\Lambda^{2-2K}}{2\pi(2K-1)}+2\eta_2\Lambda^{4-4K}
\int_{\tau_1,...,\tau_4}
\eexp{-\frac{1}{2M}[|\tau_0-\tau_2|-|\tau_0-\tau_3|+|\tau_0-\tau_4|+|\tau_2-\tau_3|
-|\tau_2-\tau_4|+|\tau_3-\tau_4|]}\nonumber\\&&
\times\left\{\left|\frac{\tau_{13}\tau_{24}}{\tau_{12}\tau_{14}\tau_{23}\tau_{34}}\right|^{2K}
-\frac{1}{|\tau_{12}\tau_{34}|^{2K}}-\frac{1}{|\tau_{14}\tau_{23}|^{2K}}\right\}
\eeq
Shifting all $\tau_i\rightarrow\tau_i+\tau_0$ eliminates $\tau_0$, as expected. One can also shift by one of the $\tau_i$ to eliminate it in the power factors and then integrate it exactly in the exponent. The most efficient shift is by $\tau_3$ so as to keep the symmetry under exchange of $\tau_2,\tau_4$, hence shift $\tau_1\rightarrow\tau_1+\tau_3,\,\tau_2\rightarrow\tau_2+\tau_3,\,\tau_4\rightarrow\tau_4+\tau_3$, furthermore shift all $\tau_i\rightarrow 2M\tau_i$ and finally $\tau_3\rightarrow -\tau_3$,
\beq{172}
&&1=\frac{8\eta\Lambda M}{2\pi(2K-1)} +2\eta_2(2M\Lambda)^{4-4K}\,I(K)\\&&I(K)=
\int_{\tau_1,\tau_2,\tau_4}\eexp{-|\tau_2|+|\tau_2-\tau_4|-|\tau_4|}
\left\{\left|\frac{\tau_{1}\tau_{24}}{\tau_{12}\tau_{14}\tau_{2}\tau_{4}}\right|^{2K}
-\frac{1}{|\tau_{12}\tau_{4}|^{2K}}-\frac{1}{|\tau_{14}\tau_{2}|^{2K}}\right\}
\int_{\tau_3}\eexp{-|\tau_3-\tau_2|+|\tau_3|-|\tau_3-\tau_4|}\nonumber
\eeq

The subsection below details how the leading form at $K=\half$ is evaluated. The critical $\eta$, using $\eta_2=\frac{\eta^2}{(2\pi)^2 4N}$, near $K=\half$, is the solution of
\beq{173}
1=\frac{4\eta\Lambda M}{2\pi(K-\half)} -\frac{16\eta^2}{4(2\pi)^2N(K-\half)^2}(2M\Lambda)^2
\eeq
which is solved by $\eta_c$
\beq{174}
\eta_c=\frac{K-\half}{M\Lambda}\cdot\frac{\pi N}{4}\left[1-\sqrt{1-\frac{4}{N}}\right]
\eeq
For large $N$ the solution is $\eta_c=\frac{K-\half}{M\Lambda}\frac{\pi}{2}(1+\frac{1}{N}+O(1/N^2))$ consistent with \eqref{e168};
Eq. \eqref{e173} has in fact a second solution with $[1+\sqrt{1-\frac{4}{N}}]$ which for large $N$ has a high slope $\eta_c\sim N$, hence is probably nonphysical.

Our main result is that the phase transition survives, at least near $K=\half$, for $N\geq 4$.
Hence the large $N$ expansion has some correspondence
with the phase diagram as found in the main text.

\subsection{C. Details of expansion near $K=\half$}

We evaluate here the critical line for the mean field transition near $K=\half$ starting from Eq. \eqref{e172}.
When integral limits are not specified they correspond to $-\infty,\infty$.
The $\tau_3$ integral needs to be done for each of the 6 orderings of $0,\tau_2,\tau_4$, however, the symmetry $\tau_2\leftrightarrow\tau_4$ and overall sign change of all $\tau_i$ yield that the 4 orderings $(0,\tau_2,\tau_4),(0,\tau_4,\tau_2),(\tau_4,\tau_2,0),(\tau_2,\tau_4,0)$ are the same and also the 2 orderings $(\tau_2,0,\tau_4),(\tau_4,0,\tau_2)$ are the same. Consider first $0<\tau_2<\tau_4$
\beq{175}
&&\int_{\tau_3}\eexp{-|\tau_3-\tau_2|+|\tau_3|-|\tau_3-\tau_4|}=\int_{-\infty}^0
\eexp{\tau_3-\tau_2-\tau_3+\tau_3-\tau_4}+\int_0^{\tau_2}\eexp{\tau_3-\tau_2+\tau_3+\tau_3-\tau_4}+
\int_{\tau_2}^{\tau_4}\eexp{-\tau_3+\tau_2+\tau_3+\tau_3-\tau_4}\nonumber\\&&+
\int_{\tau_4}^\infty \eexp{-\tau_3+\tau_2+\tau_3-\tau_3+\tau_4}=
\frac{2}{3}\eexp{-\tau_2-\tau_4}-\frac{2}{3}\eexp{2\tau_2-\tau_4}+2\eexp{\tau_2}
\eeq
The additional exponent in this range is $\eexp{-2\tau_2}$. This yields then 3 terms, $I_1,\,I_2,\,I_3$ respectively, so that
\beq{176}
I(K)=4(I_1+I_2+I_3)+2(I_4+I_5+I_6)
 \eeq
with $I_4,\,I_5,\,I_6$ are defined below for the $(\tau_2,0,\tau_4)$ ordering.
\beq{177}
I_1=\frac{2}{3}\int_{\tau_1}\int_{0<\tau_2<\tau_4}\eexp{-3\tau_2-\tau_4}
\left\{\left|\frac{\tau_{1}\tau_{24}}{\tau_{12}\tau_{14}\tau_{2}\tau_{4}}\right|^{2K}
-\frac{1}{|\tau_{12}\tau_{4}|^{2K}}-\frac{1}{|\tau_{14}\tau_{2}|^{2K}}\right\}=O(\frac{1}{2K-1})
\eeq
It turns out that all the integrals below converge at short times. The $\tau_2,\tau_4$ integrals converge, while the $\tau_1$ integral give a single $\frac{1}{2K-1}$ factor. Below we find terms that diverge more strongly at $K\rightarrow \half$, so this one can be neglected. Next is
\beq{178}
I_2=-\frac{2}{3}\int_{\tau_1}\int_{0<\tau_2<\tau_4}\eexp{-\tau_4}
\left\{\left|\frac{\tau_{1}\tau_{24}}{\tau_{12}\tau_{14}\tau_{2}\tau_{4}}\right|^{2K}
-\frac{1}{|\tau_{12}\tau_{4}|^{2K}}-\frac{1}{|\tau_{14}\tau_{2}|^{2K}}\right\}=O(\frac{1}{2K-1})
\eeq
Since $\tau_2<\tau_4$ both these integrals converge and the result can again be neglected. Next is
\beq{179}
&&I_3=2\int_{\tau_1}\int_{0<\tau_2<\tau_4}\eexp{-\tau_2}
\left\{\left|\frac{\tau_{1}\tau_{24}}{\tau_{12}\tau_{14}\tau_{2}\tau_{4}}\right|^{2K}
-\frac{1}{|\tau_{12}\tau_{4}|^{2K}}-\frac{1}{|\tau_{14}\tau_{2}|^{2K}}\right\}
\eeq
Subdivide $I_3$ in various regions of $\tau_1$ integral, i.e. $I_3=I_{31}+I_{32}+I_{33}+I_{34}$. Starting with $I_{31}$ in the range $0<\tau_2<\tau_4<\tau_1$, with the variables
$x=\frac{\tau_1-\tau_4}{\tau_2},\,y=\frac{\tau_4-\tau_2}{\tau_2}$ so that $0<x,y<\infty$ we obtain
\beq{180}
I_{31}=2\int_0^\infty d\tau_2\frac{\eexp{-\tau_2}}{\tau_2^{4K-2}}\int_0^\infty dx\int_0^\infty dy \left\{\left(\frac{y(1+x+y)}{x(x+y)(1+y)}\right)^{2K}-\frac{1}{[(x+y)(1+y)]^{2K}}
-\frac{1}{x^{2K}}\right\}\nonumber\\
\eeq
The $\tau_2$ integral for $K\rightarrow 1/2$ gives 1. For the $x,y$ integrals, for $K\rightarrow \half$ use the series
\beq{181}
(1-\alpha)^{2K}=1-2K\alpha+2K(2K-1)\sum_{n=2}^\infty \frac{(n-2)!}{n!}\alpha^n\qquad |\alpha|<1
 \eeq
 with $\alpha=\frac{x}{(x+y)(1+y)}<1$. The $n=0$ term cancels the divergent last term of $I_{31}$. The $n=1$ term can be integrated (using Mathematica) leading to
$\frac{-1}{4(K-\half)^2}+O(K-\half)^{-1}$, while the 2nd term of $I_{31}$ yields
$\frac{-1}{8(K-\half)^2}+O(K-\half)^{-1}$. Hence the leading term has
\beq{182}
I_{31}= -\frac{3}{4(K-\half)^2}+O(\frac{1}{K-\half})
\eeq

The next part of $I_3$ is $I_{32}$ with $0<\tau_2<\tau_1<\tau_4$ and with the variables $x=\frac{\tau_4-\tau_1}{\tau_2},\,y=\frac{\tau_1-\tau_2}{\tau_2}$ we obtain
\beq{183}
I_{32}=2\int_0^\infty d\tau_2\frac{\eexp{-\tau_2}}{\tau_2^{4K-2}}\int_0^\infty dx\int_0^\infty dy
\left\{\left(\frac{(1+y)(x+y)}{xy(1+x+y)}\right)^{2K}-\frac{1}{[y(1+x+y)]^{2K}}
-\frac{1}{x^{2K}}\right\}
\eeq
$I_{32}(K=\half)=0$, however the derivative, using the expansion $x^{2K}=x+x(2K-1)\ln x+O(2K-1)^2$ diverges,  hence $I_{32}$ may be discontinuous, and actually it is.
  The $y<1$ range can be shown numerically to be less divergent while the $y>1$ range can be evaluated with the expansion Eq. (\ref{e181}) and $\alpha=-\frac{x}{y(1+x+y)}$. Again only the $n=1$ term and the 2nd term of $I_{32}$ contribute to the leading divergence with
\beq{184}
I_{32}=+\frac{1}{4(K-\half)^2}+O(\frac{1}{K-\half})
\eeq

The next part is
\beq{185}
I_{33}= 2\int_{0<\tau_1<\tau_2<\tau_4}\eexp{-\tau_2}
\left\{\left|\frac{\tau_{1}\tau_{24}}{\tau_{12}\tau_{14}\tau_{2}\tau_{4}}\right|^{2K}
-\frac{1}{|\tau_{12}\tau_{4}|^{2K}}-\frac{1}{|\tau_{14}\tau_{2}|^{2K}}\right\}
\eeq
Mathematica v8.1 manages to evaluate this directly, with the result $I_{33}\sim 1/(K-\half)$, which is confirmed by expansions similar to those of $I_{31}$.
The final part of $I_3$ is $I_{34}$ with $\tau_1<0<\tau_2<\tau_4$. Defining $x=\frac{t_4}{t_2},\,y=\frac{t_4-t_1}{t_2}$ so that $1<x<y<\infty$,
\beq{186}
I_{34}=2\int_0^\infty d\tau_2\frac{\eexp{-\tau_2}}{\tau_2^{4K-2}}\int_1^\infty dy\int_1^y dx
\left\{\left(\frac{(1-x)(x-y)}{xy(1-x+y)}\right)^{2K}-\frac{1}{[x(1-x+y)]^{2K}}
-\frac{1}{y^{2K}}\right\}
\eeq
Using the expansion \eqref{e181} with $\alpha=\frac{y}{x(1-x+y)}$ where again the $n=1$ term and the 2nd term of $I_{34}$ dominate
\beq{187}
I_{34}=-\frac{3}{2(K-\half)^2}+O(1/(K-\half))
\eeq
Collecting all $I_3$ terms we obtain,
\beq{188}
I_3=-\frac{2}{(K-\half)^2}+O(1/(K-\half))
\eeq

Consider next the range $\tau_2<0<\tau_4$.
\beq{189}
&&\int_{\tau_3}\eexp{-|\tau_3-\tau_2|+|\tau_3|-|\tau_3-\tau_4|}=\int_{-\infty}^{\tau_2}
\eexp{-\tau_2+\tau_3-\tau_3-\tau_4+\tau_3}+\int_{\tau_2}^0\eexp{-\tau_3+\tau_2-\tau_3-\tau_4+\tau_3}
+\int_0^{\tau_4}\eexp{-\tau_3+\tau_2+\tau_3-\tau_4+\tau_3}\nonumber\\&&+
\int_{\tau_4}^{\infty}\eexp{-\tau_3+\tau_2+\tau_3-\tau_3+\tau_4}
=2\eexp{-\tau_4}-2\eexp{\tau_2-\tau_4}+2\eexp{\tau_2}
\eeq
The additional exponent in this range =1. This yields then 3 terms $I_4,\,I_5,\,I_6$ for the integral $I(K)$:
\beq{190}
&&I_4=2\int_{\tau_1}\int_{\tau_2<0<\tau_4}\eexp{-\tau_4}
\left\{\left|\frac{\tau_{1}\tau_{24}}{\tau_{12}\tau_{14}\tau_{2}\tau_{4}}\right|^{2K}
-\frac{1}{|\tau_{12}\tau_{4}|^{2K}}-\frac{1}{|\tau_{14}\tau_{2}|^{2K}}\right\}
\eeq
 Subdivide this into 4 ranges of $\tau_1$ so that $I_4=I_{41}+I_{42}+I_{43}+I_{44}$, considering first $I_{41}$ for $\tau_2<0<\tau_4<\tau_1$ with variables $x=\frac{\tau_1-\tau_4}{\tau_4},\,y=-\frac{\tau_2}{\tau_4}$ we obtain
\beq{191}
I_{41}=2\int_0^\infty d\tau_4\frac{\eexp{-\tau_4}}{\tau_2^{4K-2}}\int_0^\infty dx\int_0^\infty dy \left\{\left(\frac{(1+x)(1+y)}{xy(1+x+y)}\right)^{2K}-\frac{1}{(1+x+y)^{2K}}
-\frac{1}{(xy)^{2K}}\right\}
\eeq
$I_{41}(K=\half)=0$, yet as $I_{32}$ it is discontinuous.  Using \eqref{e181} with $\alpha=-\frac{1+x+y}{xy}$ and Mathematica we obtain
\beq{192}
I_{41}=+\frac{1}{2(K-\half)^2}+O(\frac{1}{K-\half})
\eeq

The next $\tau_1$ interval has
\beq{193}
I_{42}=2\int_{\tau_2<0<\tau_1<\tau_4}\eexp{-\tau_4}
\left\{\left(\frac{\tau_{1}\tau_{42}}{\tau_{12}\tau_{41}(-\tau_{2})\tau_{4}}\right)^{2K}
-\frac{1}{(\tau_{12}\tau_{4})^{2K}}-\frac{1}{(\tau_{41}(-\tau_{2}))^{2K}}\right\}
\eeq
 An expansion as in Eq. (\ref{e181}) leads to $\sim 1/(K-\half)$, hence $I_{42}$ can be neglected. The next $\tau_1$ range for $I_{43}$ has $\tau_2<\tau_1<0<\tau_4$. Defining $x=\frac{\tau_1-\tau_2}{\tau_4},\,y=-\frac{\tau_1}{\tau_4}$ yields
\beq{194}
I_{43}=2\int_0^\infty d\tau_4\frac{\eexp{-\tau_4}}{\tau_2^{4K-2}}\int_0^\infty dx\int_0^\infty dy \left\{\left(\frac{(y(1+x+y)}{x(1+y)(x+y)}\right)^{2K}-\frac{1}{(x)^{2K}}
-\frac{1}{((1+y)(x+y))^{2K}}\right\}
\eeq
 Using \eqref{e181} with $\alpha=\frac{x}{(1+y)(x+y)}<1$ to cancel the divergent 2nd term of $I_{43}$ leads to
\beq{195}
I_{43}=\frac{-3}{4(K-\half)^2}+O(1/(K-\half))
\eeq

The last interval of $I_4$ is
\beq{196}
I_{44}=2\int_{\tau_1<\tau_2<0<\tau_4}\eexp{-\tau_4}
\left\{\left(\frac{(-\tau_{1})\tau_{42}}{\tau_{21}\tau_{41}(-\tau_{2})\tau_{4}}\right)^{2K}
-\frac{1}{(\tau_{21}\tau_{4})^{2K}}-\frac{1}{(\tau_{41}(-\tau_{2}))^{2K}}\right\}
\eeq
$I_{44}(K=1/2)=0$, however it is also discontinuous. Using variables $x=\frac{\tau_2-\tau_1}{\tau_4},\,y=\frac{\tau_2}{\tau_4}$
\beq{197}
I_{44}=2\int_0^\infty d\tau_4\frac{\eexp{-\tau_4}}{\tau_2^{4K-2}}\int_0^\infty dx\int_0^\infty dy \left\{\left(\frac{(x+y)(1+y)}{xy(1+x+y)}\right)^{2K}-\frac{1}{(xy)^{2K}}-\frac{1}{(y(1+x+y))^{2K}}
\right\}\nonumber\\
\eeq
and with Mathematica we find
\beq{198}
I_{44}=\frac{1}{4(K-\half)^2}+O(\frac{1}{K-\half})
\eeq
Collecting all terms of $I_4$
\beq{199}
I_{4}=0+O(1/(K-\half))
\eeq

The next term is
\beq{200}
I_5=-2\int_{\tau_1}\int_{\tau_2<0<\tau_4}\eexp{\tau_2-\tau_4}
\left\{\left|\frac{\tau_{1}\tau_{24}}{\tau_{12}\tau_{14}\tau_{2}\tau_{4}}\right|^{2K}
-\frac{1}{|\tau_{12}\tau_{4}|^{2K}}-\frac{1}{|\tau_{14}\tau_{2}|^{2K}}\right\}=O(\frac{1}{2K-1})
\eeq
since both $\tau_2,\tau_4$ exponentially converge, only $\tau_1$ gives $\sim 1/(2K-1)$. Next one is
\beq{201}
I_6=2\int_{\tau_1}\int_{\tau_2<0<\tau_4}\eexp{\tau_2}
\left\{\left|\frac{\tau_{1}\tau_{24}}{\tau_{12}\tau_{14}\tau_{2}\tau_{4}}\right|^{2K}
-\frac{1}{|\tau_{12}\tau_{4}|^{2K}}-\frac{1}{|\tau_{14}\tau_{2}|^{2K}}\right\}=I_4
\eeq
by $\tau_2\rightarrow -\tau_4$, $\tau_4\rightarrow -\tau_2$, $\tau_1\rightarrow -\tau_1$. Finally, our integral $I(K)$ in Eq. (\ref{e172}) is
\beq{202}
I(K)=4(I_1+I_2+I_3)+2(I_4+I_5+I_6)=-\frac{8}{(K-\half)^2}+O(1/(K-\half))
\eeq
This leads to the equation for the critical $\eta_c$, Eq. (\ref{e174}).\\

\end{widetext}


\begin{thebibliography}{99}
\bibitem{schirotzek} A. Schirotzek, C.-H. Wu, A. Sommer, and M. W. Zwierlein, Phys. Rev. Lett. {\bf 102}, 230402 (2009)
\bibitem{nascimbene} S. Nascimb\'ene, N. Navon, K. J. Jiang, L. Tarruell, M. Teichmann, J. McKeever, F. Chevy, and C. Salomon, Phys. Rev. Lett. {\bf 103}, 170402 (2009)
\bibitem{gadway} B. Gadway, D. Pertot, R. Reimann, and D. Schneble, Phys. Rev. Lett. {\bf 105}, 045303 2010).
\bibitem{spethmann} N. Spethmann, F. Kindermann, S. John, C. Weber, D. Meschede, and A. Widera, Phys. Rev. Lett. {\bf 109}, 235301 (2012)
\bibitem{zipkes} C. Zipkes, S. Palzer, C. Sias, and M. Ko¨hl, Nature {\bf 464}, 388 (2010)
 \bibitem{schmid} S. Schmid, A. H\"arter, and J. H. Denschlag, Phys. Rev. Lett. {\bf 105}, 133202 2010)
 \bibitem{chikkatur} A. P. Chikkatur, A. Go¨rlitz, D. M. Stamper-Kurn, S. Inouye, S. Gupta, and W. Ketterle, Phys. Rev. Lett. {\bf 85}, 483 (2000)
 \bibitem{palzer} S. Palzer, C. Zipkes, C. Sias, and M. Ko¨hl, Phys. Rev. Lett. {\bf 103}, 150601 (2009)
\bibitem{catani} J. Catani, G. Lamporesi, D. Naik, M. Gring, M. Inguscio, F. Minardi, A. Kantian, and T. Giamarchi, Phys. Rev. A{\bf 85}, 023623 (2012)
\bibitem{fukuhara} T. Fukuhara, A. Kantian, M. Endres, M. Cheneau, P. Schau{\ss}, S. Hild, D. Bellem, U. Schollw\"ock, T. Giamarchi, C. Gross, I. Bloch, and S. Kuhr, Nature Phys. {\bf 9}, 235 (2013)
\bibitem{meinert} F. Meinert, M. Knap, E. Kirilov, K. Jag-Lauber, M. B. Zvonarev, E. Demler and H.-C. N\"agerl, Science {\bf 356}, 945 (2017)
    \bibitem{zvonarev}  M.  B.  Zvonarev,  V.  V.  Cheianov,  and  T.  Giamarchi,
Phys. Rev. Lett. {\bf 99}, 240404 (2007)
\bibitem{schechter} M.  Schecter,  D.  Gangardt,  and  A.  Kamenev, Ann. Phys. (NY) {\bf 327}, 639 (2012)
\bibitem{massel}  F. Massel, A. Kantian, A. J. Daley, T. Giamarchi, and P. T\"orm\"a,
New J. Phys. {\bf 15}, 045018 (2013)
\bibitem{burovski} E. Burovski, V. Cheianov, O. Gamayun, and O. Lychkovskiy,
Phys. Rev. A{\bf 89}, 041601 (2014)
\bibitem{gamayun}  O. Gamayun, O. Lychkovskiy, and V. Cheianov,
Phys. Rev. E{\bf 90}, 032132 (2014)
\bibitem{kantian}  A.  Kantian,  U.  Schollw\"ock,  and  T.  Giamarchi, Phys. Rev. Lett. {\bf 113}, 070601 (2014)
\bibitem{andraschko} F. Andraschko and J. Sirker,
Phys. Rev. B{\bf 91}, 235132 (2015)
\bibitem{doggen}  E. V. H. Doggen, S. Peotta, P. T\"orm\"a, and J. J. Kinnunen,
Phys.  Rev.  A{\bf 92}, 032705 (2015)
\bibitem{visuri} A.-M.  Visuri,  T.  Giamarchi,  and  P.  T\"orm\"a,
Phys.Rev.B{\bf 93}, 125110 (2016)
\bibitem{stornaiuolo}  D. Stornaiuolo, S. Gariglio, N.J.G. Couto, A. Fete, A.D. Caviglia, G. Seyfarth, D. Jaccard, A.F. Morpurgo, and J.-M. Triscone, Appl. Phys. Lett. 101, 222601 (2012).
 \bibitem{bouchiat} A. Kasumov, M. Kociak, M. Ferrier, R. Deblock, S. Gu\'eron, B. Reulet, I. Khodos, O. St\'ephan and H. Bouchiat, Phys. Rev. B{\bf 68}, 214521 (2003).
    \bibitem{altland} A. Altland and R. Egger, Phys. Rev. Lett. {\bf 110}, 196401 (2013)
\bibitem{sela} K. Michaeli, L. A. Landau, E. Sela and L. Fu, Phys. Rev. B{\bf 96}, 205403 (2017)
\bibitem{mora} L. Herviou, K. Le Hur and C. Mora, Phys. Rev. B{\bf 94}, 235102 (2016)
\bibitem{marcus} S. M. Albrecht, A. P. Higginbotham, M. Madsen, F. Kuemmeth, T. S. Jespersen, J. Nygård, P. Krogstrup and C. M. Marcus, Nature {\bf 531}, 206 (2016)
\bibitem{kouwenhoven} S. Gazibegovic et al., Nature {\bf 548}, 434 (2017)
\bibitem{bh} B. Horovitz, T. Giamarchi and P. Le Doussal, Phys. Rev. Lett. {\bf 111}, 115302 (2013).
\bibitem{flensberg} K. Flensberg, Phys. Rev. Lett. {\bf 81}, 184 (1998).
\bibitem{nazarov}  Y. V. Nazarov and D. V. Averin, Phys. Rev. Lett. {\bf 81}, 653 (1998).
\bibitem{klesse}  R.  Klesse and  A.  Stern,  Phys.  Rev.  B {\bf 62},  16912  (2000)
\bibitem{ponomarenko} V.V.  Ponomarenko,  D.V.  Averin,  Phys.  Rev.  Lett. {\bf 85}, 4928 (2000)
\bibitem{komnik} A. Komnik and R. Egger, Eur. Phys. J. B{\bf 19}, 271, (2001)
\bibitem{gao} B. Gao, A. Komnik, R. Egger, D. C. Glattli and A. Bachtold, Phys. Rev. Lett. {\bf 92}, 216804 (2004)
\bibitem{yi1} H. Yi and C. L. Kane, Phys. Rev. B{\bf 57}, R5579 (1998).
\bibitem{yi2} H. Yi, Phys. Rev. B{\bf 65}, 195101 (2002).
\bibitem{giamarchi} T. Giamarchi {\it Quantum Physics in One Dimension}, International series of monographs on physics, Vol. {\bf 121}, (Oxford University Press, Oxford, UK, 2004).
\bibitem{footnote} For any correlation function here we use the shorthand notation $\langle B^i_\omega B^j_{-\omega}\rangle$ to denote $\lim_{T\rightarrow 0} \{T\langle B^i_{\omega_n} B^j_{-\omega_n}\rangle\}$,
where the complete $T=0$ correlation in continuum frequencies is $\langle B^i_\omega B^j_{\omega'}\rangle'=2 \pi \delta(\omega+\omega')\langle B^i_\omega B^j_{-\omega}\rangle$.
\bibitem{SM} See Supplementary material below.
\bibitem{kane} C. L. Kane and M. P. A. Fisher, Phys. Rev. Lett. {\bf 68}, 1220 (1992); Phys. Rev. B{\bf 46}, 15233 (1992).
\bibitem{fateev} V. Fateev, S. Lukyanov, A. Zamolodchikov and A. Zamolodchikov, Phys. Lett. B {\bf 406}, 83 (1997)
\bibitem{tarucha} S. Tarucha, T. Honda, and T. Saku, Solid State Commun. {\bf 94}, 413 (1995)
\bibitem{yacobi} A. Yacoby, H. L. Stormer, N. S. Wingreen, L. N. Pfeiffer, K. W. Baldwin, and K. W. West, Phys. Rev. Lett. {\bf 77}, 4612 (1996)
\bibitem{safi} I. Safi and H. J. Schulz, Phys. Rev. B{\bf 52}, R17040 (1995).
\bibitem{maslov} D. L.  Maslov  and  M.  Stone,  Phys.  Rev.  B{\bf 52},  R5539 (1995).
\bibitem{ponomarenko2} V. V. Ponomarenko, Phys. Rev. B{\bf 52}, R8666 (1995).
\bibitem{chou} Y-Z. Chou, A. Levchenko and M. S. Foster, Phys. Rev. Lett. {\bf 115}, 186404 (2015).
\bibitem{lercher} A. D. Lercher, T. Takekoshi, M. Debatin, B. Schuster, R. Rameshan, F. Ferlaino, R. Grimm and H.-C. N\"agerl, Eur. Phys. J. D{\bf 65}, 3 (2011).
\bibitem{ilzhofer}  P. Ilzh\"ofer, G. Durastante, A. Patscheider, A. Trautmann, M. J. Mark, and F. Ferlaino, Phys. Rev. A{\bf 97}, 023633 (2018)
    \bibitem{kogut} J. Kogut, Rev. Mod. Phys. {\bf 51}, 700 (1978)
\bibitem{ohta} T. Ohta, Prog. Theor. Phys. {\bf 60}, 968 (1978); T. Ohta and D. Jasnow, Phys. Rev. B{\bf 20}, 139 (1979)
\bibitem{knops} H. J. F. Knops and L. W. J. den Ouden, Physica A {\bf 103}, 579 (1980)
\bibitem{maslov2} D. L. Maslov, arXiv:cond-mat/0506035 (2005), see chapter 7.
\end{thebibliography}
\end{document}